 \documentclass [12pt,a4paper      ]{article}
\usepackage{times}

\DeclareFontFamily{OT1}{times}{}
\DeclareFontShape {OT1}{times}{m }{n }{ <-> ptmr }{}
\DeclareFontShape {OT1}{times}{bx}{n }{ <-> ptmb }{}
\DeclareFontShape {OT1}{times}{m }{it}{ <-> ptmri}{}
\DeclareFontShape {OT1}{times}{bx}{it}{ <-> ptmbi}{}
\usepackage{amsmath}
\usepackage{amsfonts}
\usepackage{amssymb}
\usepackage{latexsym}
\newcommand{\cl}{C \kern -0.1em \ell} 

\newcommand{\CON}{\overline}          

\newcommand{\Scal}{\mathbb{S}}        
\newcommand{\Vect}{\mathbb{V}}        

\newcommand{\VEC}{\vec{\kern +.1em[}} 
\newcommand{\TOR}{\vec{\kern +.2em]}} 
\newcommand{\BRA}{\langle\kern -.2em\langle} 
\newcommand{\KET}{\rangle\kern -.2em\rangle} 

\newcommand{\A}{(\hspace{.5mm})} 

\newcommand{\REV}{\sim}               

\begin{document}

\title{\bf\vspace{-2.5cm} LANCZOS - EINSTEIN - PETIAU:\\
                          From Dirac's equation to\\ nonlinear wave 
                          mechanics\footnote{Published in W.R.\ Davis, et al.,
                          Cornelius Lanczos Collected Published Papers With Commentaries
                          (North Carolina State University, Raleigh, 1998)
                          Vol.\ III, p.~2-1248 to 2-1277.} }

\author{{\bf Andre Gsponer and Jean-Pierre Hurni}\\
\emph{Independent Scientific Research Institute}\\ 
\emph{ Box 30, CH-1211 Geneva-12, Switzerland}\\
e-mail: isri@vtx.ch\\}

\date{Version ISRI-94-03.8 ~~ \today}

\maketitle

\vspace{-1.0cm}

\begin{abstract}

In 1929 Lanczos showed how to derive Dirac's equation from a more fundamental system that predicted that spin $\frac{1}{2}$ particles should come in pairs. Today, these pairs can unambiguously be interpreted as isospin doublets. From the same fundamental equation, Lanczos derived also the correct form of the wave equation of massive spin 1 particles that would be rediscovered in 1936 by Proca.

   Lanczos's fundamental system was put in Lagrangian form and generalized in 1933 by Einstein and Mayer, who used the semivector instead of the quaternion formalism. Although they not did study all possible solutions, Einstein and Mayer showed that the doublets consisted of particles with different mass and charge. In fact, there are two main classes of doublets: proton/neutron and electron/neutrino pairs.

    In trying to use Proca's equation for the electromagnetic field of the electron, Lanczos reached the conclusion that the mass term could not be a constant, but had to be a function of space-time. On very general grounds, he then proposed that the elementary solutions of a fundamental nonlinear field theory should be eigensolutions.

    In complete independence of Lanczos, G\'erard Petiau discovered in 1957 a nonlinear generalization of quantum mechanics which is very close to Lanczos ideas. In such a theory, the Hamiltonian scales with the 4th power of the proper frequency. Postulating a fundamental length equal to $\alpha r_e$, with $r_e$ the classical radius of the electron, we find that the Lanczos-Einstein-Petiau model is applicable to the problem of the mass of the quarks and leptons. The compatibility of the existence of this fundamental length with quantum theory and special relativity limits the number of quarks to five and the number of massive leptons to three.

\end{abstract}

\newpage

\noindent{\Large {\bf Contents}}
\vspace{1\baselineskip}

\renewcommand{\labelitemi}{}
\begin{itemize}

\item \ref{introd:0}. Introduction
\item \ref{threep:0}. Three problems
\item \ref{forint:0}. Formulating and interpreting Dirac's equation 
\item \ref{biquat:0}. Biquaternions and the residual arbitrariness in the definition\\
                      \rule{0mm}{1mm}~~~~of the quaternion product. 
\item \ref{sperel:0}. Special relativity and quaternions 
\item \ref{maxdir:0}. From Maxwell's equation to Dirac-Lanczos's equation 
\item \ref{funiso:0}. Lanczos fundamental equation and the isospin doubling of\\
                      \rule{0mm}{1mm}~~~~Dirac's equation
\item \ref{mincou:0}. Minimum coupling. Two-component electron theory
\item \ref{einmay:0}. Einstein-Mayer-Lanczos: proton/neutron and electron/neutrino doublets\\
                      \rule{0mm}{1mm}~~~~~in the same Lagrangian.
\item \ref{unifie:0}. Einstein-Mayer-Lanczos: towards a unified picture of fundamental\\
                      \rule{0mm}{1mm}~~~~~~~interactions? 
\item \ref{proca:0}.  Lanczos's discovery of Proca's equation 
\item \ref{spin01:0}. Quaternionic derivation of spin 0 and spin 1 wave equations
\item \ref{monopo:0}. Maxwell's equation and the absence of magnetic monopoles
\item \ref{energ:0}.  The energy-momentum tensor and the stability of matter
\item \ref{nonlin:0}. The principles of a nonlinear field theory of matter 
\item \ref{barut:0}.  Petiau's nonlinear wave mechanics and Barut's mass formula
\item \ref{length:0}. The fundamental length and the elementary particle mass spectrum
\item \ref{conclu:0}. Conclusion: The advent of nonlinear quantum theory
\item \ref{append:0}. Appendix: Comparison of Dirac-Lanczos's and Dirac's equations

\end{itemize}

\centerline{\emph{\underline{Preliminary remarks}}}

\noindent\emph{This commentary was written in 1994 and published in 1998 in the \emph{Cornelius Lanczos Collected Published Papers With Commentaries} \cite[Vol.~III, p.~2-1248 to~2-1277]{DAVIS1998-}.  The first part, Secs.~1 to 15, is a comprehensive appreciation of the papers Lanczos wrote in 1929 on Dirac's equation and relativistic quantum mechanics \cite{LANCZ1929B,LANCZ1929C,LANCZ1929D,LANCZ1930A}; the second part, Secs.~15 to 17, is a short digression, inspired by Lanczos, on the possible relevance of more recent contributions by P\'etiau \cite{PETIA1965-,PETIA1982-} and Barut \cite{BARUT1979-,GSPON1996-} to the problem of the mass spectrum of elementary particles.
~\\
~\\
\indent This corrected and updated version has been posted on {\tt ~arXiv.org~} on August 4, 2005, as a tribute to Sir William Rowan Hamilton, on the occasion of his 200th birthday.}

~\\                          \centerline{* * *}


\section{Introduction}
\label{introd:0}

In 1929, Lanczos was with Einstein in Berlin. At last, he was working with the great man to whom, ten years before, he dedicated his dissertation: a quaternionic field theory of classical electrodynamics. But Lanczos's mind was not only on general relativity. Indeed, he was about to publish a series of three papers on Dirac's equation \cite{LANCZ1929B,LANCZ1929C,LANCZ1929D}.

   The year before, Dirac had discovered his relativistic wave-equation for the electron. Because of its success in explaining both the electron spin and the fine structure of atomic energy levels, all physicists immediately recognized the paramount importance of Dirac's discovery. Moreover, the simplicity of the arguments leading to its derivation was at once taken as the proof of its absolute {\it fundamental} character, an opinion that still prevails today. Lanczos had a different interpretation.  For him Dirac's equation was a {\it derived} one: he saw that Dirac's theory was a natural development of his quaternionic field theory of 1919 \cite{LANCZ1919-}.

   With considerable assurance, and in his characteristic didactic style, Lanczos passionately puts down his derivation of Dirac's equation. He produces a very modern article, which over sixty years later still contains a number of ideas which remain at the forefront of fundamental theory. He displays his  masterful command of quaternionic algebraic techniques and, with a very careful choice of notations and minimal use of redundant mathematical jargon, he ends up with a physics paper so different in style from those of his time that one could think that he wrote it for the future.

   The exact reasons why these articles were ignored by the vast majority of his contemporaries, and the reasons why Lanczos himself abandoned quaternions and never returned to quaternionic field theory for the rest of his life, could be the subject of research for some historian of science. Already on October 25th, 1929, at the lecture he gave at the Physical Society of Berlin, he did not pronounce the word `quaternion' \cite{LANCZ1930A}. Later \cite{LANCZ1931A}, in his famous 1931 paper where he discusses the fundamental issues raised by general relativity, quantum theory and the unsolved `problem of matter,' he returned to his interpretation of matter as singularities in an otherwise regular field, and once again compared this idea with the concept of eigensolutions which has proved to be so successful in quantum mechanics. However, he will not quote any of the four papers he wrote in 1929.

   Lanczos will briefly refer to his three quaternion articles of 1929 only twice, once in 1932, in his first article in English \cite{LANCZ1932B}, and in 1933 in his new derivation of Dirac's equation \cite{LANCZ1933A}. In this derivation, Dirac's equation is found to be the equivalent of Hamilton's first order equations in a problem where a second order relativistic potential equation is in correspondence with the second order Euler-Lagrange equations. Moreover, as in 1929, Dirac's equation is not found in isolation. There is a doubling in the number of solutions, which from four in Dirac's theory (two for spin and two for particle/antiparticle) increases to eight, a feature that we can today interpret as isospin.\footnote{However, in his second article of 1933 Lanczos finds no doubling of the Dirac equation. This is due to the fact that the four-potential is taken as real. Allowing this potential to become complex, two independent Dirac equations would be found again.}

   The isospin partner of the proton --- the neutron --- was discovered in 1932, at a time when the existence of the neutrino was already postulated. Thus, one may wonder why nobody ever thought of using Lanczos's doubling to explain the existence of these particles. The fact is that history took a different course, and it now remains for a historian of science to explain why.

   Another mystery is that Lanczos discovered the correct equation for a massive spin one particle, including its conserved current and energy-momentum tensor, seven years before it was rediscovered by Proca in 1936. Moreover, the whole series of Lanczos's articles is a remarkable discussion of the fundamental problems of matter, fields, and the origin of mass, most of which is still pertinent today. In particular, his intuition led him to foresee the possibility of a nonlinear theory, exactly of the kind that Petiau will develop almost thirty years later, and which is likely to give an answer to problem of the mass spectrum of elementary particles.

\section{Three problems}
\label{threep:0}

   Of course, as in any new development, there were problems in Lanczos quaternionic field theory, and Lanczos himself was very much aware of them. The sad thing is that essentially nobody, including some of the most influential physicists of the time, tried to follow and possibly improve on Lanczos's work. Among those rare contemporary physicists who took interest in Lanczos's approach, only three seem to have found it worthwhile to refer to Lanczos's quaternionic papers: Franz Sauter in Germany, Gustave Juvet in Switzerland, and Alexandre Proca in France. We will come back to them later.

   In short, with hindsight, we can easily find the main problems in Lanczos's papers. These problems are of three types.

 First, there are problems of physical interpretation because Lanczos's equation is much more general than Dirac's. For instance, apart from the difficulties due to the prediction of antiparticles, Dirac's equation was very successful because it perfectly suited electrons. The trouble with Lanczos's fundamental system (from which Dirac's equation can be derived as a special case) is that it allows for spin~$\frac{1}{2}$ solutions (such as the electron), as well as for spin 0, 1 and $\frac{3}{2}$.\footnote{In this paper we are not going to consider explicitly the spin $\frac{3}{2}$ solutions. However, the main point is that since a spinor is composed of $2S+1$ components, one biquaternion is enough to describe a spin $\frac{3}{2}$ spinor ($S=\frac{3}{2}$). See \cite{GSPON2003-}.} Obviously, just like anybody at the time, Lanczos was completely unaware of the richness of the elementary particles spectrum: he thought that his equation described electrons, and that the `doubling' was possibly related to the existence of protons.

   The second problem has to do with spatial reversal. It seems that Lanczos was not familiar with the idea that covariance with respect to spatial reversal had also to be included in order to have full relativistic invariance. For this reason, a rather substantial part of all articles is devoted to giving indirect arguments to justify the suppression of unphysical terms (such as `magnetic currents'), or to discussing complicated equations in which contributions of different parities are mixed. 

   The third problem is more technical and is connected with the previous one. It has to do with the general question of how to write strictly covariant expressions in terms of quaternions, including spatial reversal. Given that Lanczos knew a lot about quaternions, in general, it follows that Lanczos would have appreciated the solution to this problem if someone had solved it at the time. Unfortunately, he had to do his best, going back and forth between quaternions and tensors, in order to circumvent the difficulty, the precise nature of which he was not really aware.

   Because of these problems, it is not easy to read and comment on Lanczos's papers in detail. We have therefore decided to write our commentary rather independently of them, although we shall try to follow the order in which Lanczos introduced the main concepts as much as possible. Since Lanczos's papers are very dense and detailed, we shall concentrate on the more fundamental parts. Where required, we shall also correct some of his formulas on the basis of our hindsight. In effect, while some of his formulas or wording need updating, it is remarkable that Lanczos always remained extremely close to the truth as we know it today. Obviously, taking this approach we are at the risk of crediting Lanczos with discoveries that in actual fact he did not consciously make.

\section{Formulating and interpreting Dirac's equation}
\label{forint:0}

In the introduction to his first paper \cite{LANCZ1929B}, Lanczos starts by recalling the two conditions which led Dirac to formulate his equation: it had to be a linear differential system of first order, and the iterated equation had to give the `relativistic Schr\"odinger wave-equation,' i.e., what we today call the `Klein-Gordon equation.' The first condition is sufficient for having a deterministic evolution equation, and the second one insures the relativistic invariance of the result. Thus, with just two axioms, Dirac had a solution to the problem of the relativistic electron, which, as stressed by Lanczos, has the advantage of being amenable to general `operator methods,' for which one does not have to worry about the difficulties of normal tensor analysis.

   Today, Dirac's equation is a cornerstone of physics. Just like Schr\"odinger's equation in the non-relativistic domain, it is the focal point around which the highly successful formalism of contemporary relativistic quantum electrodynamics is built. Hence, despite the fact that many big problems are still unsolved (i.e., mass, charge, infinities, etc.), it is almost heretical to discuss the possibility that Dirac's equation is not the most fundamental one.  Moreover, it is futile to think that by recasting Dirac's theory in another formalism, one may get a serious clue for a better theory of the electron and other fundamental particles.

   In fact, Lanczos was not the first to worry about the algebraic structure and tensor-analytical relations of Dirac's equation. He quotes Madelung\footnote{Erwin Madelung was the first to show that Dirac's equations are somehow equivalent to Maxwell's with feedback.  Lanczos was Madelung's assistant in Frankfurt from 1924 until Fall 1928 when he moved for one year to Berlin to become Einstein's first assistant.} \cite{MADEL1929-} and Darwin \cite{DARWI1928-} but today we could cite many authors because this question has become the source of an endless stream of studies. In effect, while for the majority of physicists the answer would seem to have been the invention of `spinor calculus' by van der Waerden, Laporte, and Uhlenbeck \cite{VANDER1929-}, or simply the pragmatic approach of using `Dirac matrices,' for others the question is still open. In fact, the contemporary trend would seem to be a return to Lanczos, i.e., to the reformulation of Dirac's equation in the framework of Clifford algebras \cite{RIESZ1957-} of which quaternions are precisely the simplest non-trivial example.

   It is therefore interesting to note that those few who quoted Lanczos's work in 1930, were Sauter who was part of a number of young German physicists (mainly around Sommerfeld) using Clifford algebras to study Dirac's theory \cite{SAUTE1930A}, and 
Juvet\footnote{Gustave Juvet (1896-1936), translated in French (in collaboration with R. Leroy) Hermann Weyl's book ``Space, Time, Matter'' (A.\ Blanchard, Paris, 1922). Juvet's PhD thesis, ``Sur une \'equation aux d\'eriv\'ees fonctionnelles partielles et une g\'en\'eralisation du theor\`eme de Jacobi'' (Blanchard, Paris, 1926) will be used by Paul Weiss in 1938 (Proc.\ Roy.\ Soc.\ {\bf A169} page 104). For a biography and bibliography, see Actes Soc.\ Helv.\ Sci.\ Nat.\ {\bf 117} (1936) 422--426.} 
 who was leading a group of Swiss physicists who attempted to phrase special relativity, Maxwell's and Dirac's theories, also in the framework of Clifford algebras \cite{JUVET1930-}. But both efforts faded out, possibly because of the unfavorable times in Germany, and the unfortunate death of Juvet in Switzerland. Similar work on Dirac's theory took place in France where, independently, Proca\footnote{Alexandre Proca (1897-1955) translated in French (in collaboration with J.\ Ullmo) Dirac's ``Principles of Quantum Mechanics''' (Press Universitaire de France, 1931). For a biography and bibliography of Proca, see G.A. Proca, ``Alexandre Proca. Oeuvre scientifique publi\'ee'' (Georges A.\ Proca, Paris, 1988) ISBN-2-9502854-0-6.} tried to introduce without success the Clifford formalism \cite{PROCA1930C}.\footnote{In these papers Proca follows J.A.\ Schouten,  Proc. Royal Acad. Amsterdam {\bf 32} (1929) 105--108, in calling the elements of the Dirac-Clifford algebra `quadri-quaternions.'  Other independent attempts were made in England by A.S.\ Eddington, Proc.\ Roy.\ Soc.\ {\bf A 121} (1928) 524--542, and G.\ Temple, Proc.\ Roy.\ Soc.\  {\bf A 127} (1930) 339-348; in Russia by D.\ Iwanenko and K.\ Nikolski, Zeits.\ f.\ Phys.\ {\bf 63} (1930) 129--137; in Belgium by Th.\ de Donder and Y.\ Dupont, Bull.\ Acad.\ Roy.\ de Belg.\ Cl.\ Sc.\ {\bf 16} (1930) 1092--1097, {\bf 18} (1932) 596--602, {\bf 19} (1933) 472--478, and {\bf 19} (1933) 593--598; and possibly in other countries.}  The `gamma-matrices' and spinor formalisms became therefore the standard notations for formulating Dirac's theory.

   But Lanczos's motivation was not to find the ideal formulation of Dirac's equation.  He wanted to understand Dirac's theory in terms of ordinary tensor calculus, and, even more, to show its connections with his own previous work on Maxwell's theory. Indeed, Lanczos's starting point is the observation, traditionally attributed to Hermann Weyl \cite{WEYL-1928-}, that Dirac's equation in the case of a vanishing rest mass is identical to Maxwell's equation in vacuum (provided Maxwell's field is allowed to become a four-component complex field). Since this equation has a straightforward tensor interpretation, and is particularly simple when expressed in the quaternion formalism, Lanczos immediately saw a new perspective for understanding Dirac's equation, and, in his own words, to discover a ``{\it via regia} to grasp the internal substance of Dirac's equation'' \cite[p.~449]{LANCZ1929B}.

\section{Biquaternions and the residual arbitrariness
         in the definition of the quaternion product}
\label{biquat:0}

Quaternion formulations in physics can be seen as the most simple holistic four-dimensional formalism, in which complicated physical variables are treated as single entities that are not just abstract symbols, but hypercomplex numbers which can be directly used for explicit calculations. In general, the quadruplets will consist of complex numbers, hence the name biquaternions.\footnote{The algebras of real and complex quaternions are denoted by $\mathbb{H}$ and $\mathbb{B}$, respectively.} For example, a point in Minkowski's space-time can be represented by the biquaternion $X=[it,\vec x\,]$. There are three fundamental involutions: quaternion conjugation, represented by a `bar,' i.e., the operator $\CON{\A}$, which changes the sign of the vector part of the biquaternion; imaginary conjugation denoted by a star $\A^*$; and biconjugation which is the combination of the two previous involutions: $\A^+ = \CON{\A}^{*}$.

   With quaternions, physical formulas are generally very compact and there is kind of a mystery that we call {\it Hamilton's conjecture} \cite{GSPON1993-}, as to why so many fundamental laws of physics can be neatly expressed in quaternion form.\footnote{Ref.~\cite{GSPON1993-} can be useful as a concise introduction to quaternions and their use in physics.} But, as with any formalism, there are a number of specific problems that have to be carefully addressed in order to have a consistent set of working rules. This is particularly the case with discrete operations, such as imaginary conjugation or, as we have already mentioned, spatial reversal. To understand the origin of these problems, it is useful to digress and to return to Sir William Rowan Hamilton's {\it Elements of Quaternions}, and more precisely to a note added by Charles Jasper Joly in 1898, at the end of section ten \cite[p.~162]{HAMIL1891-}. This will allow us to recall some definitions and fix our notation.

   Starting from the set of quadruplets of real or complex numbers, the quaternion algebra is obtained by requiring their product to be associative, and the division to be feasible always, except possibly in some singular cases.\footnote{In $\mathbb{H}$ the division is always possible. In $\mathbb{B}$, however, division by singular quaternions (also called null quaternions), for which $S\CON{S}=0$, is not possible.} Then, writing two quadruplets $A$ and $B$ as $[a,\vec a\,]$ and $[b,\vec b\,]$, and using contemporary vector notations, their product has the following explicit form 
\begin{equation}\label{1}
 [a,\vec a\,] [b,\vec b\,] = [ ab + p \; \vec a \cdot \vec b , a \vec b + \vec a b + q \; \vec a \times \vec b \,] . 
\tag{1}
\end{equation}
The two constants $p$ and $q$ are related by the equation
\begin{equation}\label{2}
 q^2 + p^3 = 0 ,
\tag{2}
\end{equation}
which shows that there is some residual arbitrariness when defining the product of two quadruplets. For instance, taking $p=-1$, $q$ can be equal to either $+1$ or $-1$. On the other hand, taking $p=+1$, $q$ may be $+i$ or $-i$.

   Obviously, in the present context, the purpose of quaternions is to provide an explicit representation of Minkowski's space-time. Taking four real numbers as our fundamental quadruplets, we can examine the consequences of specific choices of $p$ and $q$ in this context.

   The arbitrariness in the sign of $p$ is connected with the signature of the metric. In effect, since the square of the norm of a quaternion $A$ is by definition its product by its conjugated quaternion, we have $|A|^2 = A\CON{A} = a^2+p{\vec a}^2$. Thus, as long as imaginary conjugation is not given a physical interpretation, the sign of $p$ is in principle immaterial, because one can always multiply the vector or scalar part of all quaternions by $i$ in order to get the desired signature.

   The arbitrariness in the sign of $q$ is due to the non-commutativity of the quaternion product. Indeed, changing the order of the factors $A$ and $B$ is equivalent to changing the sign of $q$. We call the involution associated with the changing of this sign {it ordinal conjugation}\footnote{Or \emph{order reversal}, not to be confused with Clifford algebra `reversion.'} and designate this operation by the symbol $\A^\REV$. Obviously, since $q$ is the sign associated with the vector product, there is a close connection between ordinal conjugation and space inversion. However, contrary to the case of $p$, there appears to be no overall covariant criterion to decide for the sign of $q$, or to compensate for an arbitrary choice. Therefore, in accord with the principle of relativity, one has to make sure that fundamental physical entities are `ordinal covariant,'\footnote{Or `order covariant.'} i.e., that they do not arbitrarily depend on the sign of $q$. In particular, we will later show how to use ordinal conjugation to construct quaternion tensors of definite parities.

   This is the technical problem to which we were referring in Sec.~2. If Lanczos had its solution, he would have much more easily discovered Proca's equation.

   Returning to the choice of $p$, we see that $p=+1$ gives Hamilton's algebra, and $p=-1$ the Pauli algebra. As is well known for real components, Hamiltons' algebra is a closed 4-element algebra, while Pauli's algebra is not. Since complex numbers are to be used as components in the general case, it may seem that the choice between Hamilton's and Pauli's algebra is without consequences. In fact, while the tradition of quantum mechanics has historically developed from Pauli's choice not to use Hamilton's quaternions for his non-relativistic theory of spin, but his own set of `matrices' instead,\footnote{In fact, the first correct treatment of non-relativistic spin is by P.\ Jordan, Zeits.\ f.\ Phys., {\bf 44} (1927) 473--480. In this paper, Jordan uses quaternions explicitly. It was published a few weeks before Pauli's paper. To have a perspective on how somebody of the stature of Sommerfeld felt about Pauli's notation version quaternions, it is interesting to read his comment {\it in} A. Sommerfeld, ``Atombau und Spektrallinien - Wellenmechanischer Erg\"anzungsband'' (F. Vieweg und Sohn, Braunschweig, 1929) p.~317. See also W.\ Franz, Ref.~\cite{SAUTE1930A}.} it turns out that Lanczos's choice to keep following Hamilton was the good one. The reason is that, when working with Hamilton's biquaternions, $i$ is unique, always explicit, and independent of the rules of the algebra.\footnote{In the Pauli and Dirac formalisms there are two \emph{different} imaginary units: one in the respective $\sigma_\mu$ or $\gamma_\mu$ matrices, and one in the complex numbers multiplying them.  This leads to special rules with regards to complex conjugation, and to mathematical difficulties which are often not fully appreciated.  See, e.g., A.J.\ Macfarlane, J.\ Math.\ Phys.\ {\bf 3} (1962) 1116--1129, where it is shown that ``the Pauli matrix four-vector $\sigma_\mu$ does not by itself provide an adequate basis for the discussion of two-dimensional unimodular matrices.''} In a global field theory which encompasses classical and quantum theory, complex numbers and imaginary conjugation can then always be given a definite and consistent physical interpretation.

\section{Special relativity and quaternions}
\label{sperel:0}

In paragraphs 2, 3 and 4 of his first paper \cite{LANCZ1929B}, Lanczos recalls the main features of the quaternion formulation of special relativity that he had already developed in his dissertation \cite{LANCZ1919-}. In this formulation, contrary to tensor or spinor calculus, there are no indices and all basic physical entities are represented by simple biquaternionic expressions. The {\it variance}\footnote{For brevity, we would like to characterize the transformation properties of a given geometric object by the French term `variance.'} is then specified at the same time as each entity is defined. For example, the most basic tensor, the {\it quadrivector}, is defined by the following variance:\footnote {In tensor calculus, this variance corresponds to the fundamental expression $V{_\mu^\prime} = L{^\nu_\mu}V_\nu$ where $L{^\nu_\mu}$ could be represented by a $4 \times 4$ matrix. Just like in spinor formalism, quaternions allow this matrix to be broken down into two pieces, which in quaternions correspond to a pre-multiplying and a post-multiplying factors.}
\begin{equation}\label{3}
 V^\prime = {\mathcal{L}} V {\mathcal{L}}^+ 
\tag{3}
\end{equation}
where ${\mathcal{L}}$ is a biquaternion of norm unity, i.e.,
\begin{equation}\label{4}
    \CON{\mathcal{L}}{\mathcal{L}} = {\mathcal{L}}^+ {\mathcal{L}}^* = 1 
\tag{4}
\end{equation}
The fundamental Minkowski invariant is then simply
\begin{equation}\label{5}
   \CON{V}V = \CON{V^\prime}V^\prime = \text{invariant scalar} 
\tag{5}
\end{equation}
which shows that ${\mathcal{L}}$ represents a general Lorentz transformation. Since any normalized biquaternion can be resolved in a product ${\mathcal{L}}={\mathcal{R}}{\mathcal{B}}$, where ${\mathcal{R}}={\mathcal{R}}^*$ is real and ${\mathcal{B}}={\mathcal{B}}^+ $ is bireal, the general Lorentz transformation consists of the product of a rotation ${\mathcal{R}}$ by a boost ${\mathcal{B}}$. Multiplying quadrivectors, and alternating their variance by making use of quaternion conjugation, one obtains more complicated tensors. For example, taking two quadrivectors $V_1$ and $V_2$, the tensor
\begin{equation}\label{6}
 F = \CON{V_1}{V_2} 
\tag{6}
\end{equation}
has a {\it hexavector} variance:
\begin{equation}\label{7}
 F^\prime = {\mathcal{L}}^* F {\mathcal{L}}^+ 
\tag{7}
\end{equation}
 Because of \eqref{4}, $f=\Scal[F]$, the scalar part of the hexavector $F$ is an {\it invariant}. On the other hand, $\vec f = \Vect[F]$, the vector part of $F$, has a hexavector variance. 

   The interesting thing, which explains the power of the quaternionic tensor calculus, is that any time some 4-dimensional quantity is introduced with a given variance, the only possible different variance it may obtain is the result of operating with one of the three fundamental quaternionic involutions: $\CON{\A}$, $\A^*$, or $\A^+ $. This gives four possibilities that are analogous to the four basic variances of tensor/spinor calculus: contra- or co-variant, dotted or undotted indices. Hence, with quaternions, the conjugation symbols play the role of the position and type of the index in tensor calculus.  The use of these symbols has the effect of keeping the variance of all quantities explicit, while at the same time keeping the notation compact.

   At this point, it becomes clear that the quadrivector or hexavector variances are not the most simple ones. For instance, one can have a {\it spinor} variance:
\begin{equation}\label{8}
   F^\prime = {\mathcal{L}} F 
\tag{8}
\end{equation}
where ${\mathcal{L}}$ could be replaced by $\CON{\mathcal{L}}$, ${\mathcal{L}}^*$ or ${\mathcal{L}}^+ $, or act on the other side, giving a total of eight possible spinor variances. It is immediately seen that the norm of such spinors is again a relativistic invariant, and that nothing prevents us of using spinors as fundamental physical quantities. Unfortunately, Lanczos did not fully realize this possibility. Throughout his three papers, he will insist on keeping only `quadrivector' or `hexavector' variances as fundamental ones.\footnote{The same was already done in his doctoral dissertation, where in chapter~4 he already noted the mathematical possibility of the spinor variance \cite[Eq.~(4.8)]{HURNI2004-}.}

   A special type of Lorentz transformation which was not considered by Lanczos is that of spatial reversal. For this transformation, in the case of tensors constructed by multiplying quadrivectors, there is a simple connection with ordinal invariance. For example,
\begin{equation}\label{9}
 V = A \CON{B}C + C \CON{B}A 
\tag{9}
\end{equation}
 where $A$, $B$ and $C$ are ordinal invariant quadrivectors, is invariant under ordinal conjugation. This implies that $V$ cannot depend on the sign of $q$ (see Eq.~\eqref{1}). $V$ is thus a polar quadrivector (odd parity). On the other hand,
\begin{equation}\label{10}
 W = A \CON B C - C \CON B A 
\tag{10}
\end{equation}
changes sign with ordinal conjugation. $W$ is thus necessarily an axial quadrivector (even parity). Hence, for relativistic covariant entities of this type, symmetry or antisymmetry under ordinal conjugation is equivalent to being of odd or even parity under an improper Lorentz transformation. In the case of spinors, similar considerations can be made but the result is not a simple rule as with tensors constructed by multiplying quadrivectors.

\section{From Maxwell's equation to Dirac-Lanczos's equation}
\label{maxdir:0}

In quaternions, Weyl's equation in vacuum is simply
\begin{equation}\label{11}
 \nabla F(X) = 0 
\tag{11}
\end{equation}
Here, $\nabla = [\partial_{it},\partial_{\vec x}]$ is the 4-gradient operator, assumed to transform as a quadrivector, and $F$ the field, a biquaternionic function of $X$, which reduces to a bivector (i.e., a complex 3-vector) in the case of Maxwell's equation.

   In paragraph 5 of his first paper \cite{LANCZ1929B}, Lanczos compares equation \eqref{11} for the cases in which it is interpreted as either Maxwell's or Weyl's equation. In the Maxwell case, $F$ is the electromagnetic field bivector, $\vec E+i\vec B$. The variance is then necessarily that of a hexavector because the components of the electromagnetic bivector have to transform among themselves. In Weyl's case, however, there is no such restriction: the variance is not fully determined by \eqref{7} alone. In fact, Lanczos writes \cite[Eq.~(32) p.~456]{LANCZ1929B} 
\begin{equation}\label{12}
 F^\prime = {\mathcal{L}}^* F G 
\tag{12}
\end{equation}
where $G$ may be any biquaternion. In modern language, we would say that when $F$ has a spinor variance, equation \eqref{11} is invariant under a non-Abelian gauge transformation consisting of a right-multiplication by a normed biquaternion $G$. For Lanczos, however, this is seen as a defect, or at best as a peculiar property of the wave function that has still to be explained.

   Lanczos's next step, in paragraph 6, is to consider the first order linear wave equation obtained by putting a mass term on the right-hand side of \eqref{11}. Then, writing
\begin{equation}\label{13}
 \CON\nabla D = m D^* 
\tag{13}
\end{equation}
one gets an equation that is very similar to Dirac's, except for the imaginary conjugation on the right hand side which is essential for covariance. In particular, $D$ being a biquaternion, it has four complex components, just like Dirac's bispinor. However, iterating the equation, since $\nabla^* =-\CON{\nabla}$, Lanczos finds that the Klein-Gordon equation, $\nabla\CON\nabla D = -m^2 D$, has the wrong sign. Thus, \eqref{13} corresponds to a particle with an imaginary mass. Because of the imaginary conjugation, this problem is not corrected by replacing $m$ by $im$. In fact, what is missing is a quaternion factor on the right-hand side of $D^*$. Taking for example, $i\vec e_3$ for such a factor, it is a matter of elementary algebra to verify that the equation
\begin{equation}\label{14}
 \CON\nabla D = m D^* i \vec e_3 
\tag{14}
\end{equation}
is indeed strictly equivalent to Dirac's. This equation, to be called {\it Dirac-Lanczos's equation} \cite[Eq.~(63) p.~462]{LANCZ1929B}, will later be rediscovered by a number of people, in particular by Feza 
G\"ursey\footnote{For a curriculum vitae of Feza G\"ursey (1921-1992), and for a list of publications until 1984, see "Symmetries in Particle Physics", edited by I. Bars, A. Chodos and C.-H. Tse (Plenum Press, New York, 1985).
} \cite{GURSE1950A} 
and David Hestenes\footnote{David Hestenes (1933-) has been successful in popularizing the use of Clifford algebras in the context of special relativity, Maxwell, and Dirac theories. See, in particular, "Space-Time Algebra" (Gordon and Breach, New York, 1966, 1987, 1992).  David Hestenes's father, Magnus Hestenes, was one of Lanczos's colleagues at the Institute of Numerical Analysis at the University of California at Los Angeles in the early 1950's.   A photograph of David Hestenes's mother in company of Lanczos, \emph{circa} 1951, is reproduced on page 3--463 of Ref.~\cite{DAVIS1998-}.
} \cite{HESTE1967A}. 
In fact, in one form or another, it is the basic equation in any Clifford type formulation of Dirac's theory. In this equation, the usual arbitrariness in the choice of the `Dirac matrices' corresponds to the freedom of taking any arbitrary unit vector as a factor on the right hand side.\footnote{See the Appendix for a brief comparison of some of the remarkable differences between the Dirac and the Dirac-Lanczos equations.}

   In the remaining part of paragraph 8, Lanczos calculates the bilinear covariants of Dirac's equation and some of their properties. In quaternion formalism, this is very easily done. One finds,
\begin{align*}
 \text{1 complex invariant (scalar):}                    \hskip 0.5 in
      I_{~~} &= \Scal[D\CON D]                           \label{15}\tag{15} \\
 \text{3 complex hexavectors (antisymmetric tensors):}   \hskip 0.5 in
 \vec I_n &=  \Vect[D\vec e_n\CON D]                     \label{16}\tag{16} \\
 \text{1 bireal quadrivector (current):}                 \hskip 0.5 in
      J_{~~} &= D D^+                                    \label{17}\tag{17} \\
 \text{3 bireal pseudo-quadrivectors (axial currents):}  \hskip 0.5 in
      J_n &= Di\vec e_nD^+                               \label{18}\tag{18}
\end{align*}
These bilinear tensors have a simple interpretation. For instance, the existence of a magnetic moment associated with spin corresponds to the complex bivector $\vec I_3$. This is the antisymmetric electromagnetic polarization-magnetization momentum tensor. The four bireal quadrivectors form an orthogonal Frenet frame which moves along with the particle \cite{GURSE1955-}, a property that Lanczos associates with Einstein's concept of `distant parallelism.' From equation \eqref{14}, i.e., Dirac-Lanczos's equation for an electron without an electromagnetic field, it follows that the particle current $J$, as well as the axial currents $J_1$ and $J_2$, are conserved, while (in general) the spin density axial current $J_3$ is not conserved (it is conserved, e.g., for plane waves). According to G\"ursey \cite{GURSE1955-}, it seems that $J_1$ and $J_2$ will not be noticed until de Broglie will rediscover them in 1940.

\section{Lanczos's fundamental equation and the 
        isospin doubling of Dirac's equation}
\label{funiso:0}

As we have already said, Lanczos's main interest is not in reformulating Dirac's equation with quaternions, but in showing how to derive it from a more fundamental equation. His idea is to consider a system of two coupled equations, in which case the Klein-Gordon equation is automatically satisfied. This leads him in paragraph 7 and 8 to postulate and study the properties of his {\it fundamental equation} \cite[Eq.~(54) p.~459]{LANCZ1929B}, which we write
\begin{equation}\label{19}
 \CON\nabla A = m B \hskip 1 in \nabla B = m A
\tag{19}
\end{equation}
Obviously, compared with Dirac's equation, this system has twice too many equations and variables. But, as shown by Lanczos, it is in fact equivalent to {\it two} Dirac-type equations, independent of each other {it provided} $A$ and $B$ have a spinor variance, i.e.,
\begin{equation}\label{20}
 A^\prime = {\mathcal{L}} A G_{\text N} \hskip 1 in B^\prime = {\mathcal{L}}^* B G_{\text N}
\tag{20}
\end{equation}
and $G_{\text N}$, an optional gauge transformation, is any element of a group which will be determined below. To obtain this result, Lanczos uses a rather long argument that can be summarized and somewhat generalized as follows:

   Given an arbitrary idempotent\footnote{An idempotent $\sigma = \sigma^2$ has the property that $\sigma\CON\sigma = 0$ . For definitiveness, we take $\sigma=\frac{1}{2}(1+i\vec e_3)$, where $\vec e_3$ is the third quaternion unit.} biquaternion $\sigma$, every biquaternion $Z \in \mathbb{B}$ can be written as a linear combination
\begin{equation}\label{21}
 Z = Z_1\sigma + Z_2\CON\sigma
\tag{21}
\end{equation}
where $Z_1\in \mathbb{H}$ and $Z_2\in \mathbb{H}$ are two real quaternions. In spinor language, this general theorem simply means that if we interpret a biquaternion as a (4-component)-bispinor, it can always be decomposed into a pair of independent (2-component)-spinors. Hence, for $Z$ equal to $A$ or $B$, $\sigma$ can be used as a projection operator which enables one to define two independent linear superpositions, each of which is a new bispinor:
\begin{equation}\label{22}
 D_1 = A \sigma + B^* \CON\sigma
\tag{22}
\end{equation}
and
\begin{equation}\label{23}
 D_2 = (A \CON\sigma - B^* \sigma) i\vec e_1
\tag{23}
\end{equation}
These superpositions are Lorentz covariant because of \eqref{20}, and the factor $i\vec e_1$, orthogonal to $\vec e_3$, has been introduced in order to agree with the modern isospin interpretation. It is then easily verified that both $D_1$ and $D_2$ are solutions of Dirac-Lanczos's equation \eqref{14}. The allowed gauge transformations $G_{\text N}$ are obviously those transformation which leave the superpositions (\ref{22}--\ref{23}) invariant. Since these transformations have also to leave the probability current $J$ invariant, it follows from \eqref{17} and \eqref{20} that they have to be unitary, i.e., such that $G_{\text N}{G_{\text N}}^+ = 1$. Hence, the allowed gauge transformations are elements of the group $UN(1,\mathbb{C})$ which comprises all unitary quaternions that commute with $\sigma$. Consequently, the only gauge invariant bilinear covariants are $I$, $\vec I_3$, $J$ and $J_3$, i.e., the bilinear covariants of the standard Dirac theory.

   The group $UN(1,\mathbb{C}) \simeq U(1) \otimes U(1)$, whose elements have the following explicit representation
\begin{equation}\label{24}
 G_{\text N} = \sigma \exp(-\vec e_3 \alpha) + \CON\sigma \exp(-\vec e_3 \beta) \tag{24}
\end{equation}
will be rediscovered by Nijishima \cite{NISHI1957-} in 1957. He will show that this is the largest group leaving the Dirac equation and the structure of quantum electrodynamics invariant.

   In the gamma-matrices formulation of Dirac's theory, the interpretation of Nishijima's group is not obvious. But here it is trivial. Suppose we make the spinor decomposition of $A$ and $B$ according to \eqref{21} and multiply from the right by $G_{\text N}$, the result is
\begin{equation}\label{25'}
 A G_{\text N} = A_1 \exp(-\vec e_3 \alpha) + A_2 \exp(-\vec e_3 \beta) \tag{25'}
\end{equation}
\begin{equation}\label{25''}
 B G_{\text N} = B_1 \exp(-\vec e_3 \alpha) + B_2 \exp(-\vec e_3 \beta) \tag{25''}
\end{equation}
Comparing with (\ref{22}--\ref{23}), we immediately see that while $A$ and $B$ transform with the same gauge $G_{\text N}$, $D_1$ and $D_2$ transform with $\exp(-\vec e_3 \alpha)$ and $\exp(+\vec e_3 \beta)$, respectively. In other words, if we interpret $G_{\text N}$ as a local gauge, $D_1$ and $D_2$ correspond to states with different couplings to the gauge field.

   To get closer to the modern isospin interpretation, we make the following gauge transformation:
\begin{align}\label{26}
 G_{\text N}(\mathcal{N},\gamma) &=
     \sigma\exp(-\vec e_3 \frac{\mathcal{N}+1}{2}\gamma)
   + \CON\sigma\exp(+\vec e_3 \frac{\mathcal{N}-1}{2}\gamma)   \notag \\
         &= \exp(i\frac{\mathcal{N}}{2}\gamma) \exp(-\vec e_3\frac{1}{2}\gamma)
\tag{26}
\end{align}
Hence, if $\gamma$ is the electromagnetic gauge field, and $\mathcal{N}$ the nucleon number, the electric charges of $D_1$ and $D_2$ are respectively $Q_1=\frac{\mathcal{N}+1}{2}$ and $Q_2=\frac{\mathcal{N}-1}{2}$. This is nothing but Heisenberg's formula which gives the electric charge in function of the nucleon number and the projection of isospin:
\begin{equation}\label{27}
 Q = \frac{\mathcal{N}}{2} + t_3
\tag{27}
\end{equation}
This formula was proposed soon after the discovery of the neutron, and later was generalized to hypercharged particles by Gell-Mann and Nishijima. For $\mathcal{N}=1$, $D_1$ is the proton ( $t_3=+\frac{1}{2}$ ) and $D_2$ the neutron ( $t_3=-\frac{1}{2}$). 

   The interpretation of $D_1$ and $D_2$ as an isotopic doublet was given for the first time by G\"ursey \cite{GURSE1958-} in 1957. G\"ursey also showed that the isospin raising/lowering operator is right multiplication of $A$ and $B$ by $i\vec e_1$. This interpretation, and the fact that the isospin raising/lowering is done by operating on $A$ and $B$, confirms that the fundamental fields are Lanczos's fields $A$ and $B$, and that the Dirac fields $D_1$ and $D_2$ are derived ones.

   Lanczos, of course, could not have given this interpretation. He published his papers in volume {\bf 57} of {\it Zeitschrift f\"ur Physik}, while Weyl's seminal paper on the local gauge interpretation of electrodynamics had just appeared in the previous volume \cite{WEYL-1929-}. Lanczos's interpretation was that the `phase factor' was without consequence since it would disappear in what he called a ``quantummechanical hermitian operation.'' In any case, Lanczos had the electron-proton doublet in his mind. Nevertheless, he was right to be convinced that the decomposition of his fundamental system \eqref{19} into two independent Dirac's equations was of fundamental physical importance.

\section{Minimum coupling. Two-component electron theory.}
\label{mincou:0}

Now that we have found the correct interpretation of Lanczos's doubling, it is worthwhile to have a look at Lanczos's second paper \cite[p.~476]{LANCZ1929C} in which he considers a particle in a non-vanishing electromagnetic field. As he did not use the local gauge invariance concept, Lanczos used the minimum coupling substitution instead, i.e., the replacement of $\nabla$ by $\nabla - e\phi$ in \eqref{19}, where $\phi$ is the potential of the Maxwell field. He then made the superposition \eqref{22}, once with the idempotent $\sigma=\frac{1}{2}(1+i\vec e_3)$, and a second time with the sign of $i\vec e_3$ reversed. This gave the following equations
\begin{equation}\label{28}
 \CON\nabla D_1 - e\CON\phi D_1 i\vec e_3 = + m {D_1}^* i\vec e_3
\tag{28}
\end{equation}
\begin{equation}\label{29}
 \CON\nabla {D_1}^\prime + e\CON\phi {D_1}^\prime i\vec e_3 = - m {D_1}^{\prime *} i\vec e_3
\tag{29}
\end{equation}
Hence, for these superpositions, the two Dirac fields have electrical couplings of opposite signs, something that is explicitly remarked by Lanczos \cite[p.~476]{LANCZ1929C}. But, as the right hand sides show that the masses are equal and of opposite signs, we see that the electron/proton interpretation that Lanczos was looking for is not consistent.

   In fact, if instead of Lanczos's second superposition leading to \eqref{29}, we make the proper orthogonal isospin superposition \eqref{23} --- after having made the minimum coupling substitution in \eqref{19} --- we find that both $D_1$ and $D_2$ satisfy the wave equation \eqref{28} and therefore have the {\it same} electric charge. This shows that the minimum coupling substitution is not sufficient to get the correct isospin interpretation of Lanczos's fundamental system. For this interpretation, it is necessary to make the gauge transformation \eqref{26} of the fields $A$ and $B$.

   In the same paper, Lanczos calculates the second order equation for his fundamental fields $A$ and $B$:
\begin{equation}\label{30}
 (\nabla\CON\nabla - m^2 + e^2 \phi\CON\phi) A = e (\phi\CON\nabla + \nabla\CON\phi) A
\tag{30}
\end{equation}
\begin{equation}\label{31}
 (\nabla\CON\nabla - m^2 + e^2 \phi\CON\phi) B = e (\CON\phi\nabla + \CON\nabla\phi) B
\tag{31}
\end{equation}
For comparison, let us include the result for Dirac-Lanczos's field \eqref{28} as well
\begin{equation}\label{32}
 (\nabla\CON\nabla - m^2 + e^2 \phi\CON\phi) D = e (\phi\CON\nabla + \nabla\CON\phi) D i\vec e_3
\tag{32}
\end{equation}
In this equation, the biquaternion $D$ is equivalent to a 4-component Dirac spinor. But, since a biquaternion can always be split in two parts according to the decomposition \eqref{21}, \eqref{32} can be replaced by a pair of equations of the form \eqref{30} in which $A$ is a singular quaternion, i.e., equivalent to a 2-component spinor \cite{BLATO1935-}. In this case, each of these singular second order equations is equivalent to one 2-component spinor second order equation.

   Hence, when $A$ is singular, \eqref{30} is equivalent to the relativistic Pauli equation (sometimes called the Kramer's equation \cite{KRAME1935-}) which will be reinvented by Feynman and Gell-Mann \cite{FEYNM1958-} in 1958 in order to justify the 2-component theory of weak interactions. In fact, with a proper interpretation, this second order equation is totally equivalent to the first order Dirac equation.\footnote{Already in 1933 in his new derivation of Dirac's equation \cite{LANCZ1933A}, Lanczos showed that either a first order or second order formalism can be used to describe a spin~$\frac{1}{2}$ particle.} It can therefore be used to reformulated quantum electrodynamics in 2-component spinor formalism instead of the usual 4-component one \cite{BROWN1958-}. The same applies to the singular quaternion version of \eqref{32} which leads to an elegant and compact quaternionic formulation of quantum electrodynamics \cite{GSPON1992-}. It is therefore a great compliment to Lanczos, thirty years later, to read in Feynman and Gell-Mann's paper \cite[p.~194]{FEYNM1958-}: ``... let us imagine that [Lanczos's second order equation] had been discovered first, and [Dirac's first order equation] only deduced from the later.''

\section{Einstein-Mayer-Lanczos: proton/neutron and electron/neutrino
        doublets in the same Lagrangian}
\label{einmay:0}

When he wrote his 1929 papers on Dirac's equation, Lanczos was with Einstein in Berlin. It is very plausible that Lanczos had some discussions with Einstein on his ideas about quaternions and Dirac's equation. The fact is that Einstein with Walter Mayer, his next collaborator after Lanczos's departure to the United States, will develop a new four-dimensional formalism: the {\it semivectors} \cite{EINST1932-}. This theory is rather similar to quaternions, except that the metric is hyperbolic instead of elliptic.\footnote{The close proximity of semivectors to quaternions was pointed out by J.A.\ Schouten, Zeits.\ f.\ Phys.\ {\bf 84} (1933) 92--111, and others, e.g., \cite{BLATO1935-}.} Because they originated from Einstein, they remained a parallel development to spinor calculus for a few years.

   In their first paper, submitted in 1932, Einstein and Mayer rewrote Lanczos's fundamental equation \eqref{19}, i.e., \cite[Eq.~(50) p.~539]{EINST1932-}, as well the Dirac-Lanczos equation \eqref{14}, i.e., \cite[Eq.~(51) p.~540]{EINST1932-}, in semivector form.  However, in neither case did they credit Lanczos.  Instead, after noting that the Dirac-Lanczos equation could be derived from the fundamental equation by ``specializing'' the later in such a way that it contains just one four-dimensional field and its complex conjugate, Einstein and Mayer referred to the Dirac-Lanczos equation as a ``Schr\"odinger equation.''  

   In 1933, Einstein and Mayer derived a spin~$\frac{1}{2}$ semivector field equation (in fact, a generalized form of Lanczos's fundamental equation) predicting that particles would come in doublets of different masses \cite{EINST1933A}.\footnote{In fact, Einstein and Mayer actually considered the most general Lagrangian including the generalized Dirac-Lanczos field, the Maxwell field, and gravitation.} Because of the similarity of semivectors to biquaternions, we can easily translate this model into our notation. The idea was that the most general Lagrangian for quaternionic fields, to be called the {\it Einstein-Mayer-Lanczos (EML) Lagrangian}, should have the form:
\begin{equation}\label{33}
 L = A^+ \cdot\CON\nabla AC + B^+ \cdot\nabla BC - (E^+ A^+ \cdot B + B^+ \cdot AE) + \bigl(...\bigr)^+
\tag{33}
\end{equation}
where the $C$ and $E$ fields are invariant, except possibly for spatial reversal. The field equations are then
\begin{equation}\label{34}
 \CON\nabla A C = B E^+ \hskip 1 in \nabla B C = A E
\tag{34}
\end{equation}
which reduce to \eqref{19} when $C=1$ and $E=m$. For the modern interpretation, we can put $C=1$.\footnote{In the language of particle physics, $E$ and/or $C$ correspond to some pseudo-scalar or scalar fields.  For example, $E$ can be interpreted as the \emph{Higgs field} when $C=1$.  See, e.g., \cite{GSPON2001-}.}

   Compared with Lanczos's fundamental system \eqref{19}, we see at once that \eqref{34} is no more symmetric between $A$ and $B$: parity invariance is lost. In effect, since spatial reversal replaces $\nabla$ by $\CON\nabla$ in Lanczos's equation, the parity conjugate field of $A$ is $B$, and vice versa (in spinor language, $A$ and $B$ are opposite `chirality' fields that are interchanged in a spatial reversal). This interchange is no more possible with the generalized system \eqref{34} when $E^+ \neq E$.

   Another major difference with Lanczos's fundamental system is that the second order equations for $A$ and $B$ are now eigenvalue equations for the mass. In effect, assuming $C=1$ for simplicity, and eliminating the $A$ or $B$ fields in \eqref{34}, we find the generalized Klein-Gordon equations
\begin{equation}\label{35}
 \nabla\CON\nabla A = A E E^+ \hskip 1 in \nabla\CON\nabla B = B E^+ E
\tag{35}
\end{equation}
where $m^2$ has been replaced by $EE^+ $ or $E^+ E$. Assuming a plane wave solution of the form $A=Q\exp{i(\vec{p}\cdot\vec{x}-Et)}$, we obtain the eigenvalue equation
\begin{equation}\label{36}
 m^2 Q = Q EE^+
\tag{36}
\end{equation}
and a similar equation for $B$. The important point is that although $EE^+ \neq E^+ E$ in general, the two different values of $m^2$ given by the eigenvalue equations are the same for the $A$ and $B$ fields. It is therefore always possible to isolate the isospin eigenstates by making the isospin superpositions (\ref{22}--\ref{23}).  There are two main classes of solutions.\footnote{The first systematic study of the mass spectrum of the Lanczos-Einstein-Mayer equation \eqref{34} was made, at the request of W.\ Pauli, by V.\ Bargmann, Helv. Phys. Acta {\bf 7} (1934) 57--82.}

   The first one is when E is real: $E=E^*=T$. The field equations become
\begin{equation}\label{37}
 \CON\nabla A = B \CON T \hskip 1 in \nabla B = A T
\tag{37}
\end{equation}
In this case, spatial reversal invariance is restored if $T$ is assumed to transform into $\CON T$, which implies that $T$ can be interpreted as a pseudoscalar field. Now, because $EE^+ =T\CON T=m^2$ is a scalar, there is no mass eigenvalue equation and the two isospin states are degenerate in mass. Because of that Einstein and Mayer will discard this case. This is very unfortunate, because if we make Lanczos's isospin decomposition (\ref{22}--\ref{23}), we find that \eqref{37} corresponds to a doublet of equal masses spin~$\frac{1}{2}$ particles, and that these particles interact by means of the pseudoscalar field $T$. This interpretation will be first given by G\"ursey \cite{GURSE1960-}. He will show, as far as we know completely independently of Lanczos or Einstein-Mayer, that equation \eqref{37} describes a nucleonic field, non-locally coupled to a pseudoscalar eta-pion field. In other words, G\"ursey,'s interpretation implies that \eqref{37} explains to first order the whole of low-energy nuclear physics: protons, neutrons, charge independent pion theory of nuclear forces, etc.

   The second class is when $E$ is bireal: $E=E^+ $. There is now an eigenvalue equation because, in general, $EE^+ =E^2$ is not a scalar anymore. Then, there are always two solutions of different masses. Einstein and Mayer studied this case in detail because they were expecting to find an explanation for the existence of the electron and the proton. Of course, today we know that we cannot expect to find these particles in the same multiplet because the electron is a lepton, while the proton is made out of quarks. Nevertheless, it is interesting to note that when $C\neq 1$, the Lagrangian \eqref{33} has solutions which correspond to pairs of particles with different masses and electric charges of opposite signs.

   In fact, it is possible to recover Einstein-Mayer's result in the case $C=1$, provided one notices the existence of two basic conserved currents: the probability current $J$, and the barycharge current~$K$
\begin{equation}\label{38}
 J = AA^+ + \CON{BB^+ }
\tag{38}
\end{equation}
\begin{equation}\label{39}
 K = AEA^+ + \CON{BEB^+ }
\tag{39}
\end{equation}
While $J$ has the usual interpretation, $K$ divided by the mass corresponds to the electromagnetic current as defined by Einstein-Mayer. Then, when $E\CON E > 0$ the two solutions have equal signs for the charge, and when $E\CON E < 0$ the signs are opposite, as in Einstein-Mayer's case. However, for the modern interpretation, the most interesting solutions are obtained in the singular case where $E\CON E =0$, i.e., when $E$ is an idempotent. One solution is then massive (with $K=J$), and the other one massless (with $K=0$): an electron-neutrino doublet.

   The proton-neutron and the electron-neutrino solutions have the common feature that in both cases the $E$ field is also a global gauge field. This is very important because it provides a clear distinction with the other solutions of the EML Lagrangian. Hence, in summary, by simply writing down the most trivial quaternionic generalization of the Lagrangian giving the Lanczos's fundamental equation \eqref{19}, we obtain a theory which predicts the existence of the four most important particles of matter: the proton, the neutron, the electron and the neutrino.

\section{Einstein-Mayer-Lanczos: towards a unified picture of
         matter and fundamental interactions~?}
\label{unifie:0}

In the contemporary theory of elementary particles, a fundamental concept is the idea that all interactions are mediated by local gauge fields. In the Standard Model, the gauge fields of the unified electro-weak interaction are operating in an abstract space which corresponds to the transition-current algebra of perturbation theory. Therefore, the $SU(2) \otimes U(1)$ symmetry of this model concerns a kind of {\it superstructure} which is built upon a field theory in which the Dirac equation is the most fundamental equation of matter. In such a perspective, there is no explanation for the origin and nature of the gauge fields, or for the pairing of the leptons and quarks into isospin doublets. The whole picture can be described as a very successful and effective phenomenology.

   If instead we assume that the most fundamental equation is not Dirac's but Lanczos's equation \eqref{19}, we start with a system that explains isospin in a natural way. Moreover, if we include the generalization of Einstein and Mayer, we have a possibility of understanding the origin and nature of the gauge fields \cite{GSPON1994-}. In effect, as we have already explained, and as was shown for the first time by G\"ursey in the case of pion-nucleon interactions \cite{GURSE1960-}, the `internal' symmetries are explicit and trivial in Lanczos's equation, while only space-time symmetries are explicit in Dirac's equation. In other words, in this perspective the Lanczos fields $A$ and $B$, and the gauge fields allowed by the generalized Lanczos's equation, provide a kind of {\it infrastructure} from which the physical fields and their interactions are derived. If this picture is correct, there is no abstract `isospin space' or `current-algebra space,' but a unified field theory in which everything is defined in space-time.

   Consider first the case of electromagnetic interactions. As shown in Sec.~7, the electromagnetic gauge field is combined with the nucleonic or leptonic number gauge field in the Nishijima group. This is the largest group compatible with the Lanczos-G\"ursey isospin superpositions (\ref{22}--\ref{23}). From this group we understand electromagnetic interactions, the $(+1,0)$ and $(0,-1)$ charges of the nucleons and neutrino/electron pairs, as well as the possibility of $(+\frac{2}{3}, -\frac{1}{3})$ charges for the quark states.

   The second kind of interactions which concern both the nucleons and the leptons are the weak interactions. These parity violating interactions correspond to a non-Abelian gauge transformation which is made possible by the $E$ field in the EML equation \eqref{34}. In effect, in the absence of $E$, all gauge fields have to act symmetrically on the $A$ and $B$ fields, and thus conserve parity. On the other hand, when there is an $E$ field, gauge transformations such that $A \to AG$, $B \to B$ and $E \to G^+ E$ are feasible. The corresponding gauge field has an $SU(2)$ symmetry, and the resulting phenomenology is equivalent to that of the Standard Model \cite{BJORK1979-}.

   For the leptons, the weak interaction is the only operating non-Abelian field. This limitation is due to the fact that $E$ is singular.

  For the nucleons, $E=E^*=T$ is a regular quaternionic field.  One can thus perform isotopic rotations on the $A$ and $B$ fields independently. The gauge symmetry is thus the chiral $SU(2) \otimes SU(2)$ symmetry of low-energy strong interactions. Therefore, as we have explained in Sec.~10, the $T$ field leads to the strong interactions in the form of the well-known charge-independent pion-nucleon theory. Of course, as far as a general theory of strong interactions is concerned, this theory has long been superseded by `quantum chromodynamics.' One could therefore object that such a primitive formulation would be a return to the 1960's and to the naive `sigma model' of G\"ursey and Gell-Mann-Levy \cite{GURSE1960-}. Indeed, that is the case, and provided one takes into account all the subsequent progress that has been made in the interpretation of this theory, there is no contradiction with the high energy limit where `quarks,' `confinement,' and `asymptotic freedom' become the key concepts. The contemporary problem is precisely to find a new theory which would encompass both of these low and high energy limits \cite{RHO--1994-}.

   To conclude, we note that the Einstein-Mayer generalization of Lanczos's equation is obtained by the substitution $Am \to AE$ in the EML Lagrangian. Since this substitution is not symmetrical between $A$ and $B$, the lifting of the mass degeneracy of Lanczos's original equation \eqref{19} and the possibility of weak interactions, are both linked to a maximally parity violating field.

\section{Lanczos's discovery of Proca's equation}
\label{proca:0}

In the last and longest section of his first 1929 paper \cite{LANCZ1929B}, Lanczos starts his struggle in order to give a normal tensor interpretation to his fundamental equation. His motivation is clearly that since he has been successful in deriving Dirac's equation from it, the fundamental equation itself --- once properly interpreted --- should say something important about the nature of the electron and in particular the origin of its mass. It is at this point that the difficulties we already mentioned (with physical interpretation, spatial reversal, and `ordinal' covariance) become important.

   In the case of Lanczos's fundamental equation \eqref{19}, a proper covariant interpretation requires the definition of a non-ambiguous variance for the $A$ and $B$ fields, and (at least) the construction of a conserved and ordinal invariant particle density probability current. For the first requirement, Lanczos easily finds that there is only one simple possibility (provided that space reflections are not taken into account). One field has to be a quadrivector, and the other a hexavector:
\begin{equation}\label{40}
 A^\prime = {\mathcal{L}} A{\mathcal{L}}^+ \hskip 1 in B^\prime = {\mathcal{L}}^*B{\mathcal{L}}^+
\tag{40}
\end{equation}
Under these conditions, \eqref{19} is no more symmetrical between $A$ and $B$, and $A$ can be called the `potential' and $B$ the `field.'

   The particle current must be a bireal quadrivector, and Lanczos easily identifies a candidate expression, $A^+ B+B^+ A$, which, unfortunately, is {\it not} conserved:
\begin{equation}\label{41}
 \CON\nabla \cdot (A^+ B + B^+ A) \neq 0
\tag{41}
\end{equation}

   Lanczos then enters into a long discussion, which he will continue in the following papers. He will get very close to the root of the problem and, mostly following his intuition, he will discover what he was looking for: a fully covariant relativistic equation with a traditional tensor interpretation. 

   Alexandre Proca will derive his equation in 1936 \cite{PROCA1936-}. But, most extraordinary, Proca's equation (in tensor form) is already explicitly written down in Lanczos's first paper \cite[Eqs.~(98A,B), p.~472]{LANCZ1929B} and will be repeated in the following ones \cite[p.~491]{LANCZ1929D} and \cite[p.~124]{LANCZ1930A}. Believing that he had found a new equation for the electron, Lanczos had in fact discovered Proca's equation. Strangely enough, Proca will make the same mistake: only in 1938, after taking the non-relativistic limit, he will discover that his equation does not describe an electron, but a massive spin~1 particle \cite{PROCA1938-}.\footnote{In this paper, page 63, Proca qualifies the quadrivector variance ${\mathcal{L}}\A{\mathcal{L}}^+ $ as `quaternionic,' and the spinor variance ${\mathcal{L}}\A$ as `spinorial.'}

   To obtain this remarkable result, Lanczos starts by converting his quaternion system \eqref{19} into tensor form. He then removes all the terms which are linked with `magnetic currents' and finds that the resulting system has now a conserved current.  In this process, Lanczos twice uses his physical intuition of a fundamentally {\it four}-dimensional world. First, he discards the scalar part of $B$, because Lorentz invariants (such as mass or charge) should have a clear physical interpretation (such an interpretation for $\Scal[B]$ is here not readily available). Second, he discards the `magnetic current,' because the magnetic/electric `duality' is essentially a three-dimensional concept. Finally, he uses one more heuristic guide, the idea that the resulting system should have the same number of equations as of unknowns, so that, just like Dirac's equation, the retro-coupled system is homogeneous, and therefore suitable for having `eigensolutions' of the kind that are important in quantum theory.

  To conclude, it is of interest to compare Lanczos's derivation of the field equation of a massive spin one particle with that of Proca \cite{PROCA1936-}.  In effect, it is clear that Proca's derivation, which is based on the Lagrangian formalism, is much better than Lanczos's intuitive and laborious one.  But, apart from that, the main characteristics of Proca's equation, including its current and energy-momentum tensor, are already contained in Lanczos's papers of 1929.  The interesting point, which makes us wonder about the influence Lanczos's work might have had on Proca, is that Proca published in 1932 a paper \cite{PROCA1932-} in which he refers to all three of Lanczos's articles of 1929, as well as to Lanczos's dissertation of 1919.  In this article Proca uses biquaternions and tries to explain the electron-proton mass difference by means of a modified Maxwell's equation.  Actually, Proca uses Lanczos's fundamental equation \eqref{19} with $m=0$, and twice refers to it under the name of ``Lanczos's equation.''  It is therefore not unreasonable that what is now known as the ``Proca equation'' for a massive spin~1 particle should in fact be called the ``Lanczos-Proca equation.''

\section{Quaternionic derivation of the spin~0 and spin~1
         wave equations from Lanczos's fundamental equation}
\label{spin01:0}

The successful derivation of the spin~$\frac{1}{2}$ Dirac-Lanczos equation \eqref{14} came from the implicit assumption that Lanczos's double equation \eqref{19} was the fundamental one and that the $A$ and $B$ fields had to be treated on an equal footing. This meant that $A$ and $B$ were interpreted as components of a `bispinor' and that the `Dirac field' was obtained by making an appropriate covariant superposition. Hence, in the same spirit, one should not try to give a physical interpretation to \eqref{19} directly as such, but assume other possible meaningful relativistic variances, and then construct the physical fields by making some suitable superpositions.

   The derive Proca's equation, the crucial point is that while Lanczos correctly guessed that for a proper Lorentz transformation the variances assigned in \eqref{40} were adequate, he did not realize that he had to make further assumptions in order to define physical fields with a correct variance under spatial reversal as well.

   For example, if we assume that $A$ is ordinal symmetric, i.e., that $A=A^\REV$ is an odd parity quadrivector, $A$ may indeed be a physically meaningful field. On the other hand, $B=\CON\nabla A$ is neither ordinal symmetric nor antisymmetric, and therefore has {\it no} definite parity. Leaving this problem aside temporarily, we can nevertheless use $B$ as an intermediate field in order to construct a fully covariant equation for $A$. To do that, let us rewrite Lanczos's equation \eqref{19} together with its ordinal-conjugated one, and at the same substitute $B$ for $mB$ to agree with the standard notation for integer spin fields:
\begin{equation}\label{42}
 \CON\nabla A = B \hskip 1 in \nabla B = m^2 A
\tag{42}
\end{equation}
\begin{equation}\label{43}
 A^\REV \CON\nabla = B^\REV \hskip 1 in B^\REV \nabla = m^2 A^\REV
\tag{43}
\end{equation}
Then, since $A=A^\REV$, the superposition which leads to an ordinal symmetric potential is obtained by adding the two equations on the right of \eqref{42} and \eqref{43}. We get the system
\begin{equation}\label{44}
 \CON\nabla A = B \hskip 1 in \nabla B + B^\REV \nabla = 2 m^2 A
\tag{44}
\end{equation}
which is now a fully covariant equation for a polar quadrivector field $A$. But it is still not entierly correct. The reason, now that we have taken care of parity, is that the fields in \eqref{44} are in fact mixtures of spin 0 and spin 1 states. This is because the variance \eqref{40} is such that the scalar part of $B$ is an invariant while the vector part transforms as a hexavector. Hence, the field $B=[b,\vec b\,]$ can be covariantly decomposed into its scalar and vector parts. Introducing the corresponding decomposition of the potential $A=A_0+A_1$, the system \eqref{44} can be split into a spin 0 and a spin 1 part.

   For the {\it spin zero} part, because $A=A^\REV=A_0$, we see from (\ref{42}--\ref{43}) that $b=b^\REV$. The system \eqref{44} reduces therefore to the following degenerate form of Lanczos's equation
\begin{equation}\label{45}
 \CON\nabla \cdot A_0 = b \hskip 1 in \nabla b = m^2 A_0
\tag{45}
\end{equation}

   For the {\it spin one} part, the resulting equation is just \eqref{44} with the field $B$ restricted to its vector part $\vec b$.  This gives the biquaternion form of the {\it Lanczos-Proca's equation} for a massive spin~1 particle \cite[Eq.~(98)]{LANCZ1929B}, \cite[Eq.~(21)]{PROCA1936-}:
\begin{equation}\label{46}
 \CON\nabla \wedge A_1 = \vec b \hskip 1 in \nabla\vec b + \vec b^\REV\nabla = 2 m^2 A_1
\tag{46}
\end{equation}
With this equation it is trivial to verify that the correct conserved spin 1 particle current is now
\begin{equation}\label{47}
 J = {A_1}^+ \vec b + {\vec b}^+ A_1 + \bigl(...\bigr)^\REV
\tag{47}
\end{equation}
The parentheses mean that the full expression is obtained by adding the ordinal conjugated of the first part, which has the effect of making the total ordinal invariant and divergence-less, so that Lanczos's non-zero result \eqref{41} is corrected.

   The key assumption in going from Lanczos's fundamental equation \eqref{19} to \eqref{45} and \eqref{46}, i.e., the field equations for a massive `scalar' or `vector' particle, was that $A$ had to be an ordinal symmetric vector. If we had assumed the second fully covariant possibility, i.e., that $A$ is ordinal antisymmetric, we would have got the field equations for a {\it pseudo-scalar} and respectively a {\it pseudo-vector} particle. The only difference is that in the second equation of \eqref{46}, the plus sign has to be replaced by a minus sign. The fact that all equations for particles of spin 0 and 1 are degenerate cases of Lanczos's equation has first been shown by G\"ursey in his PhD thesis \cite[p.~162]{GURSE1950A}.

   Before concluding, two comments are in order. First, to get equation \eqref{44} we have added the two equations on the right of \eqref{42} and \eqref{43}. But, to be complete, we have also to take into account the equation obtained by subtracting them:
\begin{equation}\label{48}
 \nabla B - B^\REV \nabla = 0
\tag{48}
\end{equation}
This equation corresponds to the requirement that there should be no magnetic poles or currents. However, it can easily be verified that \eqref{48} is redundant and that \eqref{46} is sufficient to eliminate the magnetic contributions. The second comment is that since we have Proca' equation \eqref{46}, we can now identify the quaternion representation of the spin 1 field anti-symmetric tensor. It is
\begin{equation}\label{49}
 2 F\A = \A\vec b + \vec b^\REV\A
\tag{49}
\end{equation}
Contrary to the bivector $\vec b$, this tensor is ordinal symmetric, and therefore defines a physical field of odd parity. 

   To summarize, we have found that, apart from its decomposition into a pair of independent spin~$\frac{1}{2}$ fields, Lanczos's fundamental equation also had two possible decompositions into a spin~0 and a spin~1 field, one for scalar and vector particles and the other for pseudo-scalar and pseudo-vector particles.

\section{Maxwell's equation and the absence of magnetic monopoles}
\label{monopo:0}

As we have said in the introduction, Lanczos's motivation for introducing his fundamental equation was the idea that it was in fact Maxwell's with feedback. Now that we have derived Proca's equation \eqref{46}, we can set $m^2=0$ to get the quaternionic field equation of a mass-less particle of spin~1. Including a source term with a current J we obtain:
\begin{equation}\label{50}
 \CON\nabla \wedge A = \vec b \hskip 1 in \nabla\vec b + \vec b^\REV\nabla = -8\pi J
\tag{50}
\end{equation}
This is the correct form of {\it Maxwell's equation} in quaternions. It explicitly implies that the source current, and thus the potential, has to be ordinal symmetric and shows that the electromagnetic field is an antisymmetric tensor. The correct quaternionic form of this tensor \eqref{49} was first given by Kilmister in 1955 \cite{KILMI1955-}.

   The usual form of Maxwell's equation \cite{CONWA1911-, LANCZ1919-} contains the electromagnetic bivector $\vec b$ instead of the electromagnetic tensor \eqref{49}:
\begin{equation}\label{51}
 \CON\nabla \wedge A = \vec b \hskip 1 in \nabla\vec b = -4\pi J
\tag{51}
\end{equation}
This equation is formally incorrect because it is not covariant under spatial reversal. However, because of \eqref{48}, it is in practice equivalent to \eqref{50}, provided that $A$ and $J$ are postulated to be ordinal invariant.

   Equations (\ref{50}--\ref{51}) are valid for charged as well as for neutral massless particles. To identify the neutral fields, we observe that, in both Proca's or Maxwell's case, whenever the potential is bireal the current \eqref{47} is zero. Electrical neutrality is thus the condition $A=A^+ $, which in turns implies $J=J^+ $.

   A very important consequence of our derivation of Maxwell's equation from Lanczos's fundamental system \eqref{19} is that it implies, by \eqref{48}, the non-existence of magnetic monopoles for odd parity vector fields.\footnote{In recent years, the consequences of the null results in the search for magnetic monopoles, and in particular the implications of an essentially asymmetric form of Maxwell's equations, has been most thoroughly analyzed by E.J.\ Post. See, for example, E.J.\ Post, Phys. Rev.\ {\bf D9} (1974) 3379--3386.} This statement is valid even though we have not required $A$ or $J$ to be bireal. Hence, the non-existence of magnetic monopoles is independent of the so-called `Rainich-' or `duality-transformation' \cite{RAINI1925-}, which is simply the statement that (\ref{50}--\ref{51}) are invariant in a transformation which amounts to multiplying all terms by the same complex number.\footnote{For an analysis of the `monopole question' using 3-dimensional vector analysis, see for example, J.D.\ Jackson, Classical Electrodynamics, (Wiley, New York, 1975) 251--253. See also, J.\ Schwinger, Science {\bf 165} (1969) 757--761; and in quaternion formalism F. G\"ursey, Rev.\ Fac.\ Sci.\ Istanbul {\bf A19} (1954) 154--160.}  In other words, for Maxwell's field, it is always possible to take $A=A^+ $ and $J=J^+ $.

\section{The energy-momentum tensor and the stability of matter}
\label{energ:0}

There is a clear difference in style between Lanczos's first papers on Dirac's equation \cite{LANCZ1929B} and the whole series which followed and culminated in the last paper he wrote in Europe before leaving for the United States \cite{LANCZ1931A}.

   The first paper can be seen as a typical textbook like expos\'e, perfectly organized in consecutive sections, were everything fits together, except for the last section in which the quaternionic formalism is abandoned and various heuristic arguments are used to derive Proca's equation. From there on, the style becomes less didactic, and Lanczos starts presenting more and more speculative ideas about the unsolved problems of matter. At the same time as he abandons the purely deductive presentation of his first paper, he becomes more self-assured, and goes on to take real risks, such in the last sentences of his third quaternionic paper, where he prophesies the ``end of quantum theory as a self-contained discipline'', and the emergence of an essentially nonlinear ``theory of matter'' \cite[p.~493]{LANCZ1929D}.

   When reading the text of an invited paper he gave at the October 25, 1929, meeting of the Physical Society of Berlin \cite{LANCZ1930A}, it is not difficult to imagine the feelings of those listeners who primarily came to learn more about the relativistic electron theory which finally succeeded in explaining the atom: Who is that man, who not long after writing down Dirac's equation, objects that, in comparison with Maxwell's theory, Dirac's theory is not a true field theory, and goes on to devote more than two third of his time presenting a very ambitious and highly speculative nonlinear theory of matter?

   Clearly, for such self-confidence, the derivation from the same fundamental system, of both Dirac's and Proca's equations, must not have been enough. Similarly, the lack of interest shown by the leading quantum physicists of the time, several of them between 5 to 10 years younger than he, cannot explain why Lanczos took such a critical standpoint as to qualify quantum mechanics as a ``more rigorous and therefore more intolerant form of positivism'' \cite[p.~104]{LANCZ1931A}. Having worked with Einstein, now on his way to the United States, why did Lanczos feel the need to write such an extraordinary testimony as his thirty-five page review paper of 1931 \cite{LANCZ1931A}? Indeed, this paper discussed the fundamental theories of physics and their problems --- and important related philosophical issues --- and it also discussed Einstein's new field theory which included the concept of `distant parallelism:'

   In our opinion, Lanczos was driven by his genius. Exactly what he will later say of Hamilton could be applied to himself: ``The genius is not free to act because he is relentlessly driven by his `daimonion,' as Socrates expressed it'' \cite[p.~140]{LANCZ1967B}.

   In Lanczos's case, his ``demon'' was his conviction that the fundamental equations of matter had to contain some kind of a feedback mechanism in order to insure stability, that in their complete form these equations had to be nonlinear so that everything is fully constrained, and that the singular central region had to be explained by the same equations.

   Let us consider the question of feedback. As we already know, Lanczos was very impressed by his fundamental equation. It was Maxwell's with feedback and it had all the properties of a good field equation. In particular, he found that it had a conserved current and, moreover, a conserved energy-momentum tensor. In quaternions, this tensor has the following simple form
\begin{equation}\label{52}
 8\pi T\A = \vec b^+ \A \vec b + m^2 A^+ \A A + \bigl(...\bigr)^\REV
\tag{52}
\end{equation}
Here, as in the current \eqref{47}, making the expression ordinal symmetric has the effect of eliminating the `magnetic' terms and making the corresponding tensor symmetric. Hence, Lanczos had done more than just put a mass term on the right hand side of Maxwell's equation: he had found essentially all the important properties that Proca will rediscover in 1936.

   To understand Lanczos's interest in Proca's equation and its energy-momentum tensor, we must remember that in his dissertation he proposed a field theory of matter in which electrons were singularities in the Maxwell field \cite{LANCZ1919-}. The first term in \eqref{52} is nothing but the so-called Maxwell tensor, i.e., the energy-momentum tensor of the Maxwell field,\footnote{The quaternionic form of Maxwell's tensor was first given by Silberstein, Phil.\ Mag.\ {\bf 25} (1913) 135--144. The remarkable thing is that if this tensor would have been adopted right away, many problems due to the improper use of the non-covariant Poynting vector could have been avoided.} so that the second term is the correction arising from the feedback. Since this second term has the form of a pressure term, Lanczos's conclusion was that the vanishing of the divergence of \eqref{52} corresponds to the stabilization of the electron's field energy-momentum distribution.

   In effect, in a field theory of matter like Lanczos's electrodynamics, there is a kind of duality principle such that the physical properties of the electron can be attributed either to the singularity (the particle aspect) or to the field (the wave aspect). Thus, although it is not explicitly stated in any of Lanczos's papers, his idea was that his new equation had to describe the electromagnetic field of the electron, while Dirac's equation was only to describe the motion of the singularity.

   Such an interpretation is perfectly sensible. We all know of the infinities in classical and quantum electrodynamics that, in both cases, are somehow due to the unstable nature of the electron. In that respect, because of the coupling of Dirac's to Maxwell's equations, the success of quantum electrodynamics can be seen as the result of a cancellation between the non-zero divergence of Maxwell' tensor for the electromagnetic field and Tetrode's tensor for the Dirac field:
\begin{equation}\label{53}
    8\pi T \A = [    \Bigl( \A \cdot \CON\nabla \Bigr) D] i\vec e_3 D^+
  - D i\vec e_3 [D^+ \Bigl( \CON\nabla \cdot \A \Bigr)  ]
                  - e\Bigl( \CON\phi   \cdot \A \Bigr)    DD^+ 
\tag{53}
\end{equation}
In fact, in a truly consistent theory, both tensors should be divergenceless. Of course, starting with the introduction of stresses of an unknown nature by Poincar\'e, there have been many attempts to modify Maxwell's tensor to produce stability. In particular, the recent attempt by Schwinger \cite{SCHWI1983-} is not without resembling Lanczos's proposal to replace Maxwell's tensor with Proca's.

   It is in his third paper that Lanczos expounds his idea: ``Instead of the Dirac current we have the energy-momentum tensor, instead of current conservation, we have energy-momentum conservation'' \cite[p.~487]{LANCZ1929D}. However, when putting numbers in, it turns out that if $m$ in \eqref{46} and \eqref{52} is the mass of the electron, everything comes out wrong: the mass of the photon is no more zero but equal to $m$ ! Hence, Lanczos makes his next bold step: $m$ is in fact not the electron mass, but a ``field function'' \cite[p.~492]{LANCZ1929D} which is everywhere practically zero, except in the immediate neighborhood of the singularity. In this perspective, he suggests a new interpretation of wave mechanics: ``the de~Broglie-Schr\"odinger wave equation with constant $m$ cannot characterize a single electron but only a large `swarm' of electrons'' \cite[p.~493]{LANCZ1929D}.

   More ideas and details will be given by Lanczos in his Berlin conference \cite{LANCZ1930A}. Summarizing his results and once again writing down Proca's equation, he insists that the mass term should in fact be a field function in order that the equation can be applied to the electron. To make this nonlinearity plausible, he rewrites Proca's equation in general tensor form and then, assuming a conformally invariant geometry, he explicitly shows that the mass term indeed becomes a function of space-time.

   More than sixty years after Lanczos suggested these ideas, we are still very far from a working nonlinear relativistic quantum theory. We may even say that Lanczos's ideas were just whishful thinking. But does it mean that Lanczos's general ideas about how to modify quantum theory were wrong? Not at all. And, in our opinion, the reason is that if we believe that at some level the laws of nature must be simple, there are only a very few ways in which we can modify the existing laws of relativistic quantum theory in order to reach a more complete theory.

\section{The principles of a nonlinear field theory of matter}
\label{nonlin:0}

When Lanczos left Europe for the United States, he was in his mid-thirties and possibly was one of the few physicists of the time with a really wide perspective over the whole of fundamental theoretical physics. He had worked in classical electrodynamics, quantum theory, atomic physics, and general relativity, and knew as much about the successes as the problems of these theories. In contrast to Dirac, who  in 1930 wrote and published ``The Principles of Quantum Mechanics'' --- which tend to imply that quantum theory is a theory essentially in its final form --- Lanczos would summarize his doubts and certainties in part I of his long paper dealing with fundamental theories and related philosophical issues \cite{LANCZ1931A}.

   According to Lanczos, what are the truly fundamental principles of a physical theory of matter?

   Lanczos's starting point is a comparison of Einstein's general relativity to classical electrodynamics. In the former we have the prototype of a closed and predictive theory were everything is determined by the field equations: the geometry of space-time, the conservation laws, the motion of the masses. In classical electrodynamics, we have a satisfactory description of the electrical phenomena in the macroscopic domain, but a theory which fails at the microscopic level as soon as one meets with the problem of singularities.

   As is well known from function theory, the behavior in the neighborhood of the singularities defines the properties of the whole function everywhere. It is therefore natural to consider the singularities as the `source' of the function, and thus to see the matter/field duality as the physical counterpart of the mathematical duality of singular versus regular points of an analytic function.

   However, at the same time as the function becomes singular at a given point, it becomes undefined and as a result one has an uncertainty: the solution is by no means unique. In the most simple cases, such as in modern quantum electrodynamics, one can circumvent this difficulty by the process of `renormalization.' But universal properties such as the definite mass of electrons or protons, or the uniqueness of their electrical charge, cannot be defined by such solutions.

   Fortunately, under favorable conditions, differential equations and systems may have solutions which are everywhere regular, without being trivial. Such solutions are called `eigensolutions' and play a fundamental role in wave mechanics. If the equations are linear, a constant factor is left undetermined, but if they are nonlinear everything is fully constrained. This is the case with general relativity, and Lanczos suggests therefore that all fundamental properties of matter should be determined by eigensolutions. The position of quantum mechanics relative to fundamental field theory is therefore similar to that of Maxwell's theory relative to the classical theory of the electron: ``Wave mechanics does not solve the problem of matter, it just gives averages over space or time'' \cite[p.~112]{LANCZ1931A}.

   In this perspective, Lanczos's hope is that a future fundamental theory will solve the problem of singularities at the same time as the problem of matter and electricity. To succeed in such a task, he puts forward a two-step program: ``Generally covariant field equations giving to space-time a definite geometrical structure. Eigensolutions to these equations, giving on the one hand the electron, on the other hand the proton'' \cite[p.~112]{LANCZ1931A}.

   Sixty years later, were do we stand? All attempts to incorporate electricity into a generalized form of general relativity have failed. And the number of `elementary particles' has vastly increased, from two in 1930, to several tens or hundreds, depending on whether the excited states are counted along with the fundamental ones.

   A key element seems to be missing. Here again we may refer to Lanczos's 1931 paper. Just after giving his `program,' and before starting his analysis of Einstein's new theory of distant parallelism, he recalls the question of the universal constants: Should the fundamental equation already contain the universal constants of nature or not? Since the velocity of light allows the measurement of time in units of length, and since general relativity has reduced gravitational interaction to a question of curvature, what is the proper unit of length for a theory of elementary particles? If we exclude the singular solutions in favor of the regular ones, the eigensolutions, what are then the meanings, if any, of the Planck length and of the classical electron radius?

   The missing element is a {\it fundamental length}. If it is the Planck length, how can we explain the gigantic difference of this length relative to the electron radius? If it is the electron radius, which has about the correct size for a theory of elementary particles, why not take a length obtained by multiplying it by a dimension-less constant such as $\alpha={e^2}/{\hbar c} \simeq {1}/{137}$ ? Moreover, what physical interpretation are we going to give to this length?

   Using all our contemporary knowledge, let us try to resume Lanczos's program. Let us do it within the framework of special relativity, for the present leaving aside the question of properly dealing with gravitation. What we are looking for is some fundamental equation which could give us an explanation for the nature and basic properties, such as the mass, of elementary particles.

   As a starting point, we observe that Lanczos's fundamental equation is extremely successful because we can derive Dirac's as well as Proca's equation from it. Moreover, Lanczos's equation contains isospin, and thus an explanation of the existence of doublets of particles which are either electrically charged or neutral.

   As a second step, we consider Einstein-Mayer's generalization as a convincing hint that assuming the mass term in Lanczos's equation to be a biquaternionic parameter, we can lift the mass degeneracy of Lanczos's equation, and thus explain the existence of electron and neutrinos, and at the same time explain the existence of strongly interacting nucleon pairs. In this last case, the $E$ field introduced by Einstein and Mayer is not just a parameter but the pseudoscalar field which corresponds to the pions.

   The third step is to systematically study the possible generalization of the Lanczos-Einstein-Mayer system in order to discover other meaningful physical interpretations. In doing this generalization, it is worthwhile using, as Lanczos did, two heuristic guides. First, {\it Lanczos's feedback} idea, i.e., the concept that fundamental equations should be closed systems. Second, also following Lanczos, {\it Hamilton's conjecture}, i.e., the idea that somehow biquaternions are an ideal formalism for writing the fundamental equations of physics \cite{GSPON1993-}.

\section{Petiau's nonlinear wave mechanics and Barut's mass formula}
\label{barut:0}

In 1965, about the time when Gell-Mann and Zweig proposed the concept of quarks, in complete independence from mainstream research and without knowing that he was in the path of Lanczos, the French physicist 
G\'erard Petiau\footnote{G\'erard Petiau (1911-1990) was one of the first students of Louis de Broglie. He published many papers on waves equations of all possible spin. He is best known for his contribution to the Kemmer-Duffin-Petiau formalism of spin 1 particles. He worked out many of the mathematical details of de Broglie's `double solution' theory of causal nonlinear quantum mechanics. His major discovery is the elliptic function generalization of wave functions, which continuously interpolate between de Broglie's and solitonic waves.} 
started studying systems of equations which may precisely give a solution to the problem of the mass of leptons and quarks \cite{PETIA1965-,PETIA1982-}. In more than fifteen papers, mostly published between 1957 and 1969, Petiau analyzed many different aspects of nonlinear wave mechanics, looked at regular as well as singular solutions, and considered the problem of how to extend the canonical formalism, the question of second quantization, etc. All of that he will do mostly following the instinct of a physicist, with everything very neatly written down and explained starting from first principles, in the pure French tradition of mathematics.

   Although Petiau was thinking in very general terms, considering complicated couplings between particles of various intrinsic spin, it is easy with quaternions to write his fundamental equation in the case of spin~$\frac{1}{2}$ particles. It essentially amounts, in the spirit of Lanczos's feedback idea, to adding a third equation to Einstein's system \eqref{34} in order to close it:
\begin{align}\label{54}
     \CON\nabla A &= B E         \notag      \\
         \nabla B &= A E           \tag{54}  \\
         \nabla E &= A \CON B    \notag
\end{align}
Here $A$ and $B$ are the fundamental Lanczos fields, and $E=E^+$ an additional Einstein-Mayer field of spin~0. In the most simple case, $E$ is a real scalar function of space-time and can be seen as replacing the mass in Lanczos's fundamental equation \eqref{19}.\footnote{The function $E^2$ corresponds to the `metric factor' $\sigma$ introduced by Lanczos in his generalization of the Proca equation for conformally flat space-times \cite[p.~127]{LANCZ1930A}.} As usual, to get the physical spin $\frac{1}{2}$ eigenstates, we have just to make the Lanczos-G\"ursey isospin superposition (\ref{22}--\ref{23}).

   Because \eqref{54} is a closed nonlinear system, the solutions are much more constrained than with the usual linear type of wave equations. For instance, the single-periodic {\it de~Broglie waves} that quantum mechanics associates with a particle become double-periodic {\it Petiau waves} \cite{PETIA1958-}. Instead of being linear combinations of $\sin (z)$ and $\cos (z)$ functions, these waves are superpositions of elliptic functions $\operatorname{sn}(z,k)$, $\operatorname{cn}(z,k)$, etc. A very appealing feature of Petiau waves is that their dependence on the modulus $k$ interpolates between pure de Broglie waves (for $k=0$) and pure solitonic waves (for $k=1$): a beautiful realization of the wave/particle duality of quantum mechanics. Moreover, both the amplitudes and the proper frequency (the $\mu$ factor appearing in the argument of $\operatorname{sn}[\mu(Et-\vec p \cdot\vec x),k]$, for example) will be quantized. This is exactly the kind of behavior Lanczos was expecting.

   The most interesting thing, however, happens when, in order to quantize the system, the Hamiltonian function is constructed. Taking, for example, $A$ as the fundamental field, Petiau \cite{PETIA1965-} showed that, in terms of the first integrals, the Hamiltonian will have a very simple form
\begin{equation}\label{55}
 H = H_0 k^2 \mu^4
\tag{55}
\end{equation}
where $k$ is the modulus of the elliptic function, $\mu$ the proper frequency, and $H_0$ some constant. The exciting thing is that the Hamiltonian, and thus the total energy in the field (i.e., for a single particle, the effective mass) scales with the fourth power of $\mu$.

   In effect, in 1979, Barut \cite{BARUT1979-} discovered a very good empirical formula for the mass of the leptons. Assuming that a quantized self-energy of magnitude $\frac{3}{2} \alpha^{-1} m_e N^4$, where $N$ = 0, 1, 2..., is a new quantum number, be added to the rest-mass of a lepton to get the next heavy lepton in the chain electron, muon, tau,... , Barut got the following expression (where $\alpha \simeq {1}/{137}$):
\begin{equation}\label{56}
 m(N)= m_e (1 + \frac{3}{2} \alpha^{-1} \sum_{n=0}^{n=N} n^4 )
\tag{56}
\end{equation}
The agreement with the data of this rather simple formula is surprisingly good, the discrepancy being of order $10^{-4}$ for the muon and $10^{-3}$ for the tau, respectively: see Ref.~\cite{P.D.G1992-} or Table~1. In order to get the masses of the quarks \cite{GSPON1996-}, it is enough to take for the mass of the lightest quark $m_u={m_e}/{7.47}$ . Again, we see in Table~1 that the agreement between the theoretical quark masses and the `observed' masses is quite good, especially for the three heavy quarks \cite{GSPON1996-}.
\begin{table}
\begin{center}
\begin{tabular}{|c|c|c|c|c|c|c|} 
\hline
 N & \multicolumn{3} {c|} {\bf lepton masses} & \multicolumn{3} {c|} {\bf quark masses \rule{0mm}{4mm}} \\
\hline
   &      & Barut's  &          &         &  Barut's  &               \\
	
   &      & formula  &   Data   &         &  formula  &     Data      \\
\hline
 0 & e    &    0.511 &    0.511 & u       &     0.068 &    0 -- 8     \\
	
 1 &$\mu$ &  105.55  &  105.66  & d       &    14.1   &    5 -- 15    \\
	
 2 &$\tau$&  1786.1  & 1784.1   & s       &   239     &  100 -- 300   \\
	
 3 &      &  10294.  &    ?     & c       &  1378     & 1300 -- 1500  \\
	
 4 &      &  37184.  &    ?     & b       &  4978     & 4700 -- 5300  \\
	
 5 &      &          &          & t       & 13766     &      ?        \\
	
 6 &      &          &          &         & 31989     &      ?        \\
\hline
\end{tabular} 

\caption[Comparison of lepton and quark masses to Barut's formula]{Comparison of lepton and quark masses in MeV/c$^2$ calculated with Barut's formula \eqref{56} to measured masses from Ref.~\cite{P.D.G1992-}. The observation of a sixth quark of mass in the range of 160'000 to 190'000 MeV/c$^2$ has been reported at the beginning of 1995 \cite{ABEABACHI}. }

\end{center}

\end{table}
%
%

   Petiau's Hamiltonian \eqref{55} shows that the energy density of the solutions to his system \eqref{54} is a constant which depends only on the two parameters $k$ and $\mu$. If we assume, just like Lanczos did in 1929 \cite{LANCZ1930A}, that the singular part of the field is contained in a small region of finite extension, the `mass' of the singularity will have the same form as \eqref{55}. Moreover, without going into the details of a specific model \cite{GSPON1996-}, we can expect --- from the general principles of quantum mechanics --- that if there is a spectrum of solutions, they will correspond to integer multiples of the proper frequency $\mu$. Hence, for fixed $k$, the mass spectrum of elementary particles will correspond to a sum of partial contributions of the form
\begin{equation}\label{57}
 m(n) = m_0 \, n^4
\tag{57}
\end{equation}
Comparing with \eqref{56}, this expression establishes the link between Petiau's nonlinear fields and Barut's formula for the mass of quark and electrons. If this is correct, what about the factor 7.47 ?

   There are two non-trivial exceptional cases for elliptic functions: the harmonic case, $k=\sin({\pi}/{4})$, and the equianharmonic case, $k=\sin({\pi}/{12})$. It is very plausible to associate the former with leptons, and the latter with quarks. Indeed, in either case, the corresponding elliptic functions exhibit several unique symmetry and scaling properties, which come from the fact that in the complex plane their poles form a modular aggregate with ${\pi}/{2}$ or ${\pi}/{3}$ symmetries. Since according to \eqref{55} the mass is proportional to $k^2$, the lepton to quark mass ratio is then equal to about 7.47.

   A final very interesting feature of Petiau's nonlinear system \eqref{54} is that it has two general classes of solutions to which correspond two plane waves limits \cite{PETIA1965-}. In the first class, the third equation decouples and we get the standard Lanczos equation in which $A$ and $B$ are the two spinors of opposite chiralities which are interchanged in a spatial reversal operation. The degenerate Petiau system is thus equivalent to the Lanczos or Lanczos-Einstein-Mayer equations from which we can derive two Dirac's equations, one for each isospin states.

   In the plane wave limit of the second class, one of the spinor fields $A$ or $B$ decouples: We are left with a {\it lone spinor} coupled to the scalar field $E$. It is then not possible to construct a parity covariant field with just one degenerate Petiau system. Hence, while the first class clearly corresponds to leptons (or more generally to particles such as the nucleons which have well defined spin~$\frac{1}{2}$ asymptotic states), we can interpret the second class as {\it quark} states. `Confinement' is then explained by the above impossibility of the second class of solutions to have Dirac-like asymptotic plane wave limits.

   In order get hadrons, i.e., particles of definite parity made out of quarks, it is then necessary to make covariant superpositions of two or three quark states. In this picture, the quark states are defined as Lanczos fields and thus do not have a definite spin or parity. Hence, contrary to the Standard Model in which the quarks are Dirac fields, there is no contradiction with Pauli's exclusion principle. However, as a conclusion to these ideas, we stress that the details of such a theory of hadrons have not yet been worked out.

\section{The fundamental length and the elementary particle mass spectrum}
\label{length:0}

It is not difficult to show that the $N^4$ power law dependence for the total energy which we have found for the Lanczos-Petiau's system \eqref{54} is a general property of all Lagrangian field theories in which there is a nonlinear term in the fourth power of the field amplitude. From a theoretical point of view, this is a very nice Lagrangian because it has many good properties. In particular, it is conformally invariant and renormalizable. From a phenomenological point of view, the $N^4$ power law dependence is also very good. In effect, such a strong power law dependence is essential in order to explain the very big difference in mass between the lightest fundamental particle, the electron ($m_ec^2 = 0.5$ MeV), and the fifth and sixth quarks ($mc^2 \approx$~5000 and $\approx$~175'000 MeV). Hence, we see that Petiau's equation is more than just a generalization of Lanczos-Einstein's system and that there is a broad theoretical and empirical basis for it, although it may be just a first approximation to the complete theory.

   If we now take a purely fundamental point of view, it is important to remark that many different theories could in principle lead to a Lagrangian with a quartic term, and that in this respect Petiau's ideas are not necessarily the true ones. In particular, many nonlinear generalization of well established theories, such as gravitation \cite{KERLI1975-} or electrodynamics \cite{BISSH1972-}, quite naturally lead to such Lagrangians. The problem with these generalizations is that the scale is then either the Planck length (in which case the nonlinear effects are very small), some fraction of the electron radius $r_e$, or, as in the present model, some unknown parameter.

   Comparing expression \eqref{57} of the particle mass to the empirical formula \eqref{56}, we find that (for the lepton spectrum) the constant $m_0$ corresponds to the energy equivalent of the muon mass, i.e., about 105 MeV = $\frac{3}{2} \alpha^{-1} m_ec^2$. This figure is very interesting because the typical excitation energies of hadrons are precisely of this order. Moreover, the mass of most elementary particles have values which are very close to integer or half-integer multiples of $\alpha^{-1} m_ec^2 = 70$ MeV \cite{NAMBU1952-}. If we assume that the universal coupling constant of all interactions (except the gravitational one) of the elementary particles is defined by the electric charge quantum $e$, we derive from the formula $r_0={e^2}/{m_0 c^2}$ the {\it fundamental length} 
\begin{equation}\label{58}
r_0 = \alpha r_e
\tag{58}
\end{equation}
where we have discarded the factor $\frac{3}{2}$ because in the present discussion we are only interested in order of magnitude considerations.

   A fundamental length (or fundamental energy), expected to become the fourth fundamental constant, has been advocated by a number of authors and in particular by Heisenberg \cite{HEISE1938-} and St\"uckelberg \cite{STUCK1943-}. The problem, as we have already said, is to find its value and its correct interpretation. For Heisenberg and St\"uckelberg, the fundamental length was estimated to be of the order of $r_e$. From the mass spectrum of the electron and quarks, and the assumption that the relevant interaction constant was the electric charge, we have derived \eqref{58}. Let us give some further evidence that this is indeed the fundamental that explains the mass spectrum of elementary particles.

   The first quality of a fundamental length is that it must characterize the threshold for some new physics. In this respect, one must remember that because $e^2$ and $\hbar c$ have the same dimensions (and thus a ratio $\alpha = {e^2}/{\hbar c}$ which is pure number) one can always associate two different lengths to a given energy or mass. Hence, since the larger of these is associated with the wave-aspect and the shorter one with the particle-aspect of a massive object, each length has to be considered in the proper context. The electron mass, for example, gives the electron radius which is the distance from a point charge below which classical electrodynamics becomes inconsistent; on the other hand, the Compton wave-length of the electron is the threshold for electron-positron pair-creation, a typical phenomenon of relativistic quantum electrodynamics (QED).

   Since QED is by far the most precise and best understood part of elementary particle theory, the threshold for `new physics' in the context of QED is ---as is well known --- the phenomenon of muon pair-production.\footnote{By `new physics' we do not mean a breakdown of QED, but phenomena such as the creation of pair of particles heavier then the electron, the properties of which are not explained by QED. For electromagnetic interactions, QED has been tested to extremely short distances, corresponding to energies of the order of $10^6$ MeV.} This precisely corresponds to the energy that led to the fundamental length \eqref{58}. Hence, QED sets an {\it upper limit} to the fundamental length $r_0$, which therefore has to be at most of the order of $r_0 = \alpha r_e$ so that pairs of particles other than electrons may be created.

   In Sec.~\ref{barut:0}, in order to derive formula \eqref{57}, we have assumed that a space-time region of finite extension was associated with the particle. The size of this region, a kind of a bag which contains the energy corresponding to its mass,\footnote{It is important to stress that this quantum mechanical `bag' does not have to correspond to the extension of some directly measurable quantity, such as the electric charge or hadronic matter distribution in the bag model of hadrons. It is more like a volume of phase-space such as is introduced in field quantization.} is of course our fundamental length $r_0$. This concept can be used independently of nonlinear wave mechanics and Petiau waves, and leads to nonlinear models that are typical of those we referred to in the beginning of the present section (for an example, see Ref.~\cite{HEISE1938-}). But, whatever the details of the model, the size of this bag has to be consistent with quantum theory and with special relativity. This will give us two {\it lower limits} to the fundamental length.

   In quantum theory, a most fundamental principle is that a wave function has to be associated with every particle. In the rest frame of the particle, it has the form $\exp(i{mc^2t}/{\hbar})$. If a finite region of space-time is also associated with the particle, compatibility requires that its size has to be smaller than the wave-length of the wave function. This means that $r_0 < {\hbar}/{mc}$, which corresponds to an approximate upper limit to the mass spectrum
\begin{equation}\label{59}
 mc^2 < \frac{\hbar c}{r_0} = \alpha^{-2} m_e c^2 \simeq 9.6 \; {\text GeV} \tag{59}
\end{equation}
A second constraint on the mass of a particle of finite extension is given by the minimum radius of a relativistic particle with spin. In effect, if we attribute an intrinsic angular momentum to a relativistic particle, its center of mass cannot be defined more precisely than the ratio of the spin angular-momentum to the particle mass \cite{MOLLE1949-}. Thus, to be compatible with this relativistic effect, i.e., in order that our lepton or quark can still be considered as a spin~$\frac{1}{2}$ point particle, we need $r_0 < {\hbar}{2mc}$, which gives an upper limit of about 5 GeV, i.e., of the same order a/s the above quantum criterion.

   Looking at Table~1, we see that there is no lepton or quark with a masselarger than $\alpha^{-2}m_e = 9.6$ GeV/$c^2$. We therefore conclude that $r_0 = \alpha r_e$ is really the fundamental length we were looking for and that, as expected by Lanczos, it is neither the electron radius or the Planck length.\footnote{We remark that these considerations, and thus the mass limit \eqref{59}, applies only to quarks and leptons, i.e., to particles of `matter,' and not necessarily to bosons such the $W$ and $Z_0$, which are quanta of effective interaction fields, and which have masses of 80 and 91 $GeV$, respectively.}

   Finally, we notice that $r_0$ is essentially the electron radius times $\alpha$, the mysterious fine structure constant of Schr\"odinger. This is one more hint that $r_0$ is indeed a fundamental length. In effect, while the electron radius can be seen as the simplest combination of elementary quantities giving a scale with the dimension of a length, the determination of alpha is now equivalent to deriving $r_0$ from some more fundamental theory, unless it is assumed that $r_0$ is truly fundamental, in which case alpha is nothing but the ratio of $r_0$ to $r_e$.\footnote{It should be mentioned that the fundamental significance of $\alpha r_e$ is also suggested by classical models of particles with spin. For example, in the model of E.\ Groschwitz, the charge of the electron is supposed to be distributed over a sphere of radius $\alpha r_e.$ The motion of the electron is then a complicated `Zitterbewegung'-like trajectory which has two interesting limiting cases: one with an effective mass of the order of the electron mass, the other with an effective mass of the order of the muon mass. See pages 357--361 {\it in} H.\ H\"oln, Ergebinisse d.\ exakten Naturwiss., Bd. {\bf XXVI} (1952) 291--382.}

   In the currently Standard Model of elementary particles, a sixth quark with a very large mass is needed for theoretical and phenomenological reasons.  The discovery of this sixth quark could therefore be interpreted as an argument against the existence of a fundamental length such as required by the present interpretation of the Barut formula.  The Standard Model, however, is basically a perturbative theory of the interactions of quark and leptons, a theory that is logically independent of our interpretation of mass quantization.  In this perspective, the conjunction of our theory of mass with one of interactions could explain the need for a very massive sixth quark, as well as the existence of very light neutrinos, which are also predicted by our model \cite{GSPON1996-}.

\section{Conclusion: The advent of nonlinear quantum theory}
\label{conclu:0}

Lanczos's idea of 1929 to derive Dirac's equation from a more fundamental one was a highly heretical concept. It implied a kind of Copernician revolution which was clearly not acceptable to the `fathers of quantum mechanics.' Moreover, the essentially theoretical reasoning of Lanczos was too far from the experimental situation of the time to be confronted with the data. The fate of Lanczos's ideas is therefore somewhat similar to those of de Broglie when he proposed in 1927 his double solution theory: nobody took serious interest in it before the concept was revived in the 1950's by the work of David Bohm on deterministic quantum theory.

   The philosophical lesson is that pure theoretical insight has much more difficulties gaining acceptance than does any kind of a more phenomenological approach. Because of the brilliant success, following World War II, of the young `quantum engineers' with renormalizable quantum electrodynamics, the pragmatic approach is possibly even more in favor now than in Lanczos's days. The best example of this is the Standard Model, a highly successful effective phenomenology, which gives numerical agreement for a vast number of phenomena, but which nevertheless does not answer most of the truly fundamental problems which are the real concern of theoretical physics.

   In the highly competitive context of contemporary research, it is sad that Lanczos never learned that he had discovered isospin in 1929. This is particularly regrettable since the final interpretation of his fundamental equation was given in 1957 by G\"ursey, who therefore deserves a lot of credit for it, although in the papers which made him famous, G\"ursey did not quote Lanczos.\footnote{It is also unfortunate that a number of researchers with an interest in quaternion techniques may have been turned away from Lanczos's work on Dirac's equation by a partially incorrect statement of A.W.\ Conway, who wrote: ``Quaternions have been used by Lanczos (1929) to discuss a different form of wave equation, but here the Dirac form is discussed, the wave function being taken as a quaternion and the four-row matrices being linear functions of a quaternion.''  See, A.W.\ Conway, \emph{Quaternion treatment of the electron wave equation}, Proc.\ Roy.\ Soc.\ {\bf A 162} (1937) 145--154.} Five years later, in 1962, Lanczos will again rediscover Dirac's equation, this time in the context of general relativity \cite{LANCZ1962A}, although without the mass term. Again he will refer to quaternions and to his contention about their special role in physics:

\begin{quote}

``The spontaneous and quite unexpected appearance of quaternion calculus and the Dirac equation (the fundamental building block of electromagnetism and of quantum theory) in the field equations of general relativity can hardly be considered as mere accidents'' \cite[p.~389]{LANCZ1962A}.
\end{quote}

   Until the end, Lanczos continued the work on fundamental theory he had started in 1919 with his dissertation and that reached a climax with his review paper of 1931.  He could never forget the theoretical insight he had gained with his quaternionic derivation of Dirac's equation, and the fact that this major contribution was ignored.

   But the contemptuous attitude of Lanczos's contemporaries is not a proof of the superiority of the dominant ideas of the time.  Neither is the imperfection of Lanczos's formulation of 1929 a sufficient reason for rejecting his ideas and excluding him from the small circle of the leading physicists who were to put the dogma of the new physics into final shape.  It is therefore easy to understand the feelings of Lanczos when he criticized the intolerant positivist attitude of the quantum physicists, or announced the end of quantum theory with the advent of a more fundamental nonlinear one \cite{LANCZ1931A}.

\newpage

\section{Appendix: Comparison of Dirac-Lanczos's and Dirac's equations}
\label{append:0}

Let us compare the Dirac-Lanczos equation \eqref{14}:
\begin{equation}\label{14'}
 \CON\nabla D = m D^* i \vec \nu .
\tag{14'}
\end{equation}
to the Dirac equation:
\begin{equation}\label{60}
 \partial_\mu \gamma_\mu \Psi = m \Psi ,
\tag{60}
\end{equation}

   Two representations of the Dirac equation such that the two quadruplets of gamma-matrices are related by $\gamma_\mu' = S\gamma_\mu S^{-1}$, where $S=S^+ \in SU(4)$ is unitary, are equivalent as the far as the physical consequences of the theory are concerned.\footnote{See, e.g., J.J.\ Sakurai, Advanced Quantum Mechanics (Addison Wesley, 1987) section 3-2.} 

  In the Dirac-Lanczos equation, the arbitrariness in the choice of the `Dirac matrices' $\gamma_\mu$ corresponds to the freedom of taking any arbitrary unit vector $\vec\nu$ as a factor on the right hand side.  Therefore, two representations of Lanczos's equation such that ${\vec\nu}' = Q\vec\nu Q^{-1}$, where $Q=Q^+ \in SU(2)$ is unitary, are equivalent. 

   Consequently, in Dirac's formalism, the arbitrariness in the representation corresponds to a 15-parameter group, and in Lanczos's formalism to a 3-parameter group.

   In Dirac's formulation the 4-complex-component electron field is taken as a $4 \times 1$ column vector $\Psi$, and the linear operators are $4 \times 4$ complex matrices.

   In Lanczos's formulation the same 4-complex-component field is a biquaternion $D \in \mathbb{B} \cong M_2(\mathbb{C}) \cong \cl_{1,2} \cong \cl_{3,0}$. The linear operators are then linear biquaternions functions of biquaternions, which are isomorphic to the algebra of $4 \times 4$ complex matrices  $M_4(\mathbb{C})  \cong M_2(\mathbb{B}) \cong \cl_{4,1}$.

   In both formulations the operator space has $4 \times 4 \times 2 = 32$ dimensions over the reals. The difference is that in the Dirac formulation the field is an abstract 4-component column vector, while in the Lanczos formulation the field is directly related to the algebraic structure of spacetime because any biquaternion $D = s + \vec{v}$ is the direct sum of a scalar $s$  and a 3-component vector $\vec{v}$.

   Lanczos's formulation is therefore more suitable than Dirac's for studying and demonstrating the ``classical'' aspects of the electron field, and for making comparisons with the Maxwell and Proca fields which are usually expressed in terms of scalars and vectors \cite{GSPON2002-}.

   Finally, in terms of Clifford algebras, the Dirac field $\Psi$ is a degenerate 8-real-component element of the 32-dimensional Clifford algebra $\cl_{4,1}$ (i.e., an element of an ideal of that algebra) while the Lanczos field $D$ is any 8-real-component element of the 8-dimensional Clifford algebra $\cl_{1,2} \cong \mathbb{B}$, which is therefore the \emph{smallest} algebra in which Dirac's electron theory can be fully expressed.

   For more details see Refs.~\cite{GSPON1993-,GSPON1994-,GSPON2001-,GSPON2002-,GSPON2003-}, where it is shown that the complex conjugation operation appearing on the right-hand side of Eq.~\eqref{14'} is characteristic of the \emph{fermionic} character of the Dirac field, in contradistinction to the Maxwell and Proca fields which are bosonic fields \cite{GSPON2002-}.

\end{document}